\documentclass[aps,pra,amssymb,onecolumn,showpacs,supersriptaddress,groupeaddress]{revtex4-1}
\usepackage{amssymb}
\usepackage{amsmath}
\usepackage{graphicx}
\usepackage{float}
\usepackage{bm}
\usepackage[flushleft]{threeparttable}
\usepackage{natbib,hyperref}
\usepackage{hyperref}

\usepackage{color}
\hypersetup{colorlinks=true, 
    linkcolor=red,          
    citecolor=magenta,        
    filecolor=gree,      
    urlcolor=blue           
}

\begin{document}

\title{Optimization of the multi-mem response of topotactic redox La$_{1/2}$Sr$_{1/2}$Mn$_{1/2}$Co$_{1/2}$O$_{3-x}$}

\author{W. Rom\'an Acevedo$^{1,2}$, M. H. Aguirre$^{3,4}$, C. Ferreyra$^{1,2}$, M.J. S\'anchez$^{1,5}$, M. Rengifo$^{1,2}$, C. A. M. van den Bosch$^6$, A. Aguadero$^6$ B. Noheda$^{7,8}$, D. Rubi$^{1,2}$}


\affiliation{$^{1}$Instituto de Nanociencia y Nanotecnolog\'{\i}a (INN), CONICET-CNEA, Argentina \\
$^{2}$ Centro At\'omico Constituyentes, Av. Gral Paz 1499 (1650), San Mart\'{\i}n, Buenos Aires, Argentina\\
$^{3}$ Instituto de Nanociencia y Materiales de Arag\'on (INMA-CSIC) and  Dpto. de F\'{\i}sica de la Materia Condensada, 
Universidad de Zaragoza \\ 
$^{4}$ Laboratorio de Microscopí\'{\i}as Avanzadas, Edificio I+D, Campus Rio Ebro C/Mariano Esquillor s/n, 50018 Zaragoza, Spain. \\ $^{5}$ Centro Atómico Bariloche and Instituto Balseiro (Universidad Nacional de Cuyo), 8400 San Carlos de Bariloche, Río Negro, Argentina  \\ $^{6}$ Department of Materials, Imperial College London, London SW7 2AZ, United Kingdom \\ $^{7}$ CogniGron - Groningen Cognitive Systems and Materials Center,
University of Groningen (RuG),            
Nijenborgh 4, 9747AG Groningen, The Netherlands  \\ $^{8}$ Zernike Institute for Advanced Materials, 
University of Groningen, Nijenborgh 4, 9747AG Groningen, The Netherlands  }

\date{\today}%


\begin{abstract}

Memristive systems emerge as strong candidates for the implementation of Resistive Random Access Memories (RRAM) and neuromorphic computing devices, as they can mimic the electrical analog behavior or biological synapses. In addition, complementary functionalities such as memcapacitance could significantly improve the performance of bio-inspired devices in key issues such as energy consumption. However, the physics of mem-systems is not fully understood so far, hampering their large-scale implementation in devices. Perovskites that undergo topotactic transitions and redox reactions show improved performance as mem-systems, compared to standard perovskites. In this paper we analyze different strategies to optimize the multi-mem behavior (memristive and memcapacitive) of topotactic redox La$_{1/2}$Sr$_{1/2}$Mn$_{1/2}$Co$_{1/2}$O$_{3-x}$ (LSMCO) films grown on Nb:SrTiO$_3$ (NSTO). We explored devices with different crystallinity (from amorphous to epitaxial LSMCO), out-of-plane orientation ((001) and (110)) and stimulated either with voltage or current pulses. We found that an optimum memory response is found for epitaxial (110) LSMCO stimulated with current pulses. Under these conditions, the system efficiently exchanges oxygen with the environment minimizing, at the same time, self-heating effects that trigger nanostructural and chemical changes which could affect the device integrity and performance. Our work contributes to pave the way for the integration of LSMCO-based devices in cross-bar arrays, in order to exploit their memristive and memcapacitive properties for the development of neuromorphic or in-memory computing devices

\end{abstract}

\maketitle

\section{Introduction}

Memristors are defined tipically as metal-insulator-metal structures able to switch their resistance between different non-volatile states \cite{saw_2008, iel_2016} and are intensively investigated nowadays due to their potential for the development of a new generation of non-volatile electronic memories, coined as Resistive Random Access Memories (RRAM). In addition, many memristive systems display analog multilevel states and are, therefore, suitable to be implemented in novel neuromorphic computing devices \cite{yu_2017}, as they can mimic the electrical behavior of biological synapses \cite{orrei2020}. Different neuromorphic capabilities such as as long-term synaptic potentiation/depression \cite{yu_2017} or spike-time-dependant-plasticity \cite{prez16} have been reported for memristive devices. In addition, memristor arrays with cross-bar architecture were experimentally implemented in simple neural networks (perceptrons) for character recognition \cite{prez15}, highlighting the potential of these kind of devices for the development of bio-inspired electronics. Memristor cross-bars were also shown to be suitable as in-memory computing devices \cite{Sun4123,Sun2378}, which could solve the CPU-memory data bottleneck present in standard computers with Von Neumann architecture.

Memristive mechanisms rely on the presence and electromigration of charged defects such as oxygen vacancies (OV) \cite{saw_2008}, ubiquitously found in transition metal oxides \cite{gun20}. The presence of these defects strongly affects the local resistivity of the material. Memristive mechanisms in oxides include the formation and disruption of OV conducting nanofilaments \cite{kwon10}, the modulation by OV of interfacial metal-insulator energy barriers \cite{saw_2008} or a redox reaction taking place when a reactive electrode such as Al or Ti is deposited on top of the insulator \cite{herpers}. Complex oxides with a perovskite structure are multifunctional materials displaying in many cases memristive behavior based on OV electromigration \cite{nian2007}. In the case of the celebrated hole-doped manganites, OV disrupt double exchange Mn-O-Mn bonds and locally increase their resistivity \cite{rozenberg_2010}. In this way, an oxidized (reduced) manganite displays a more metallic (insulating) electrical behavior. Standard memristive perovskites -including manganites- can stand slight amounts of OV, leading to different reduced states which are energetically equivalent. Upon consecutive electrical cycling, any of these reduced states can be stabilized and, therefore, the corresponding resistance state presents a significant dispersion, affecting the device reliability. A more stable memristive response was shown for perovskites such as SrCoO$_3$ \cite{acharya2017,nall2019,hung2020,mou2021}, SrFeO$_3$ \cite{nalla2019, nalla2020,kim2020} or La$_{2/3}$Sr$_{1/3}$MnO$_3$ \cite{yao20}, which are able to reversibly switch to/from a brownmillerite-like (strongly reduced) phase and are examples of the so-called topotactic redox materials. For these materials, it is possible to reversibly switch between two phases with different structure and a large difference in oxygen content (and resistivity). As both oxidized and reduced states are linked to an energy landscape with well defined minima, the electrical switching between both phases is more reproducible and controlled \cite{acharya2017,mou2021}. An important issue in topotactic redox materials-based memristors is the significant oxygen exchange between the device and the atmosphere -usually neglected in memristive perovskites\cite{nian2007,rozenberg_2010}-, related to the large difference in oxygen content between oxidized and reduced phases. In order to avoid the device structural damage upon oxygen release, and maximize the device reliability, different strategies such as tuning the out-of-plane orientation of the oxide layer \cite{acharya2017,mou2021,kim2020,nalla2019,nalla2020} or including an additional material to behave as a by-design oxygen migration channel \cite{cho2016} have been proposed. In the first case, (110) or (111) orientations favour the migration of oxygen in and out the device through planes parallel to the [001] brownmillerite axis, which displays easier anionic mobility and facilitates transport between top and bottom electrodes for those out-of-plane orientations. In the second case, the fabrication of self-assembled perovskite-Sm:CeO$_{2}$ composites allows easy oxygen drift through Sm:CeO$_{2}$ vertical columns.  

A very interesting topotactic redox perovskite is La$_{1/2}$Sr$_{1/2}$Mn$_{1/2}$Co$_{1/2}$O$_{3-x}$ (LSMCO), which displays and oxidized -more conducting- phase with $x=0$ and a rhomboedral \emph{R\={3}c} structure, as well as a reduced -more resistive- phase with ${x}=$0.62 and an orthorombic \emph{Pbnm} structure \cite{agua2011}. We have shown \cite{roman2020} that epitaxial Nb:SrTiO$_3$/LSMCO structures display a robust memristive behavior concomitant with a memcapacitive effect -reversible change in the capacitance between different non-volatile states \cite{wu2011,sala2013,salarou2014,besso15,Yang_2017,park2018,Liu_2018, boro2020,guo2020}-. The memcapacitance found in this system -C$_{HIGH}$/C$_{LOW}$ $\approx$ 900 at 10 kHz and $\approx$ 100 at 150 kHz- was the highest reported to date by a factor of $\approx$ 10 and originates at the NSTO/LSMCO interface, where a switchable p-n diode is formed \cite{roman2020}. Memcapacitance presents a high potential for the development of neuromorphic applications; for instance, it has been proposed that a memcapacitor-based neural network for image recognition can perform in a similar way that one based on memristors, but with a power consumption about 1000 times lower \cite{tran2017}. In our previous work on multi-mem LSMCO, we found that the electroforming process is accompanied by a strong oxygen release and thermal effects that affect the nanostructure of the device and decouples the active zone from the rest of the device, hampering the possibility of implementing these systems in cross-bar arrays \cite{roman2020}. In this paper, we perform a systematic study of different alternatives to limit devices damage and preserve their structural integrity, chemistry and multi-mem behavior, in order to allow their integration in multiple devices architectures such as cross-bar arrays, suitable for neuromorphic or in-memory computing. 

\section{Experimental}\label{exp}

We have grown oxidized LSCMO thin films by pulsed laser deposition by using either excimer or YAG lasers. In the latter case, the deposition was assisted with high energy electron diffraction (RHEED). The films were deposited on conducting Nb:SrTiO$_3$ (0.5 wt. \%, NSTO) substrates with out-of-plane (001) and (110) orientations. The deposition temperatures ranged between RT and 850 ºC in order to obtain films with different crystallinity. The oxygen pressure was in the range 0.040-0.085 mbar, while the fluence was set between 0.5-1 J/cm$^2$. X-ray diffraction was measured with a Panalytical Empyrean diffractometer. Top Pt electrodes were deposited by either e-beam evaporation or sputtering and shaped in circles with diameters between 45 $\mu$m and 500 $\mu$m by standard optical lithography. DC Electrical characterization was performed with a Keithley 2612 source-meter hooked to a probe-station. AC electrical measurements were performed with a LCR BK894 impedance analyzer. This included impedance spectra, recorded for frequencies between 100 Hz and  500 kHz, and capacitance-voltage curves, recorded at a frequency of 10 kHz. For the latter, the impedance analyzer was set to measure on a parallel RC circuit. High resolution Scanning Transmission Electron Microscopy with High Angular Annular Dark Field Detector (STEM-HAADF) was performed using a FEI Titan G2 microscope with probe corrector (60-300 keV). In-situ chemical analysis was performed by Energy Dispersive Spectroscopy (EDS) and Electron Energy Loss Spectrocopy (EELS). Samples for TEM were prepared by Focused Ion Beam (FIB) in a Helios 650 Dual Beam Equipment.  

\section{AS-GROWN SAMPLES CHARACTERIZATION}\label{ags}

In the first place, we describe the structural properties of as-grown epitaxial LSMCO samples deposited at 800-850 ºC on (001) and (110) NSTO. Figures 1(a) and 1(b) display X-ray Bragg-Brentano patterns corresponding to (100) and (110) films, respectively. It is found the presence of LSMCO (00h) and (hh0) reflections next to the ones corresponding to the substrate. No indication of parasitic phases is found. The epitaxial character of the films is confirmed by in-situ RHEED experiments. For instance, the inset of Figure 1(a) shows a RHEED pattern corresponding to an as-grown (001) LSMCO film, displaying a stripy pattern typically found on epitaxial films with low roughness (root-mean-square roughness below 1 nm). This flat morphology is confirmed by atomic force microscopy (AFM) experiments, not shown here. The inset of Figure 1(b) shows an X-ray reflectivity experiment performed on a LSMCO (110) sample, where it can be observed the presence of non-damped oscillations which, after modelling the spectrum with GEN X software, indicates a LSMCO thickness of 16.3 nm and a surface roughness of 0.4 nm, in agreement with the morphologies infered from RHEED and AFM analysis. Figure 1(c) displays a high resolution STEM-HAADF cross-section corresponding to an LSMCO (001) film, evidencing the high structural quality of the film, with a well defined and coherent NSTO/LSMCO interface. Figure 1(d) shows EELS chemical maps of the structure, which together with EDS quantifications (not shown here) indicate that the stochiometry of the LSMCO layer corresponds to the oxidized perovskite phase.

\begin{figure}[h!]
\centering
\includegraphics[scale=0.6]{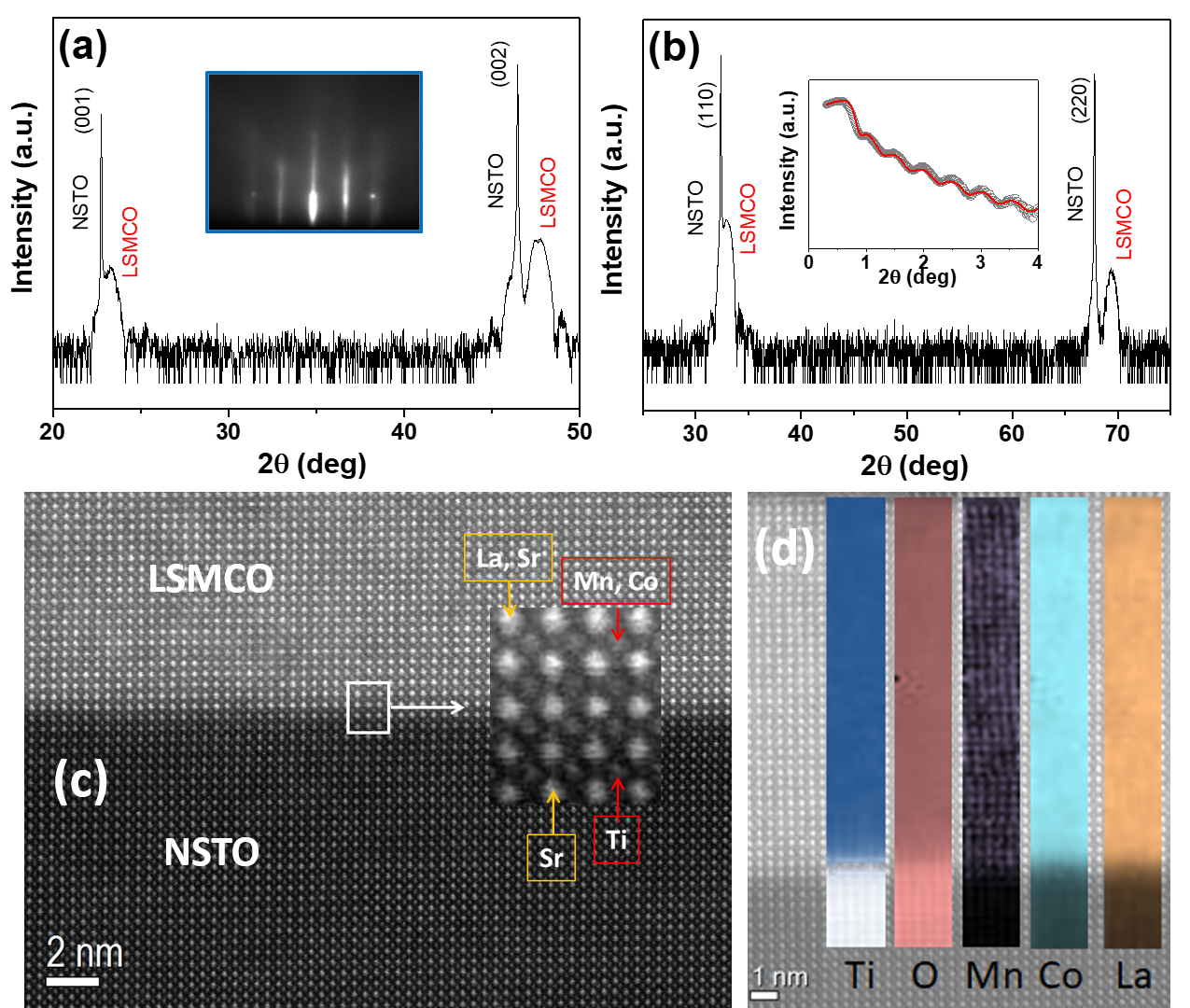}
\caption{(a) X-ray diffraction pattern corresponding to an epitaxial LSMCO thin films grown on (001) NSTO. The inset shows the RHEED pattern measured in-situ after finishing the growth process; (b) X-ray diffraction pattern corresponding to an epitaxial LSMCO thin films grown on (110) NSTO. The inset shows an X-ray reflectivity measurement (black symbols) and fitting (red line) performed on the same sample; (c) STEM-HAADF cross section corresponding to an epitaxial (001) NSTO/LSMCO film. The inset shows a zoom of the NSTO/LSMCO interface where cations have been labeled; (d) EELS maps corresponding to Ti, O, Mn, Co and La elements. Homogeneous cation distributions are seen.}
\label{fig1}
\end{figure}

Next, we describe the structure and chemistry of LSMCO films grown at low temperatures. Figure 2(a) shows a STEM-HAADF cross section corresponding to a NSTO/LSMCO film grown at 200 ºC. As evidenced by the Fast Fourier Transforms (FFT, displayed in panels (b) and (c) of Figure 2) performed on the squared zones displayed in Figure 2(a), corresponding to the LSCMO film and the NSTO substrate, respectively, LSMCO grows amorphous on single-crystalline NSTO. EDS line scans shown in panel (e) -going from the LSMCO/Pt interface to the NSTO substrate, as shown in panel (d)- indicate that the stochiometry of the amorphous LSMCO layer is, within the error of the technique, consistent with the one corresponding to the oxidized LSMCO phase.

\section{CHARACTERIZATION OF VOLTAGE-CONTROLLED DEVICES}\label{ags}

We start by reviewing the electrical response of an 16.5 nm thick epitaxial LSMCO (001) film stimulated with \emph{voltage} pulses. We refer to this sample as LSMCO1. The NSTO substrate was grounded and the electrical stimuli was applied to the top Pt electrode. The device virgin resistance was $\approx$ 1 M$\Omega$ and a forming process, consisting on a sequence of -7 V pulses that reduces the device resistance to $\approx$ 5x10$^2$ $\Omega$, was necessary to initialize the device. After forming, we recorded simultaneously dynamic I-V curves and hysteresis switching loops (HSL) that track the evolution of the device remanent resistance. For the dynamic I-V, we applied a sequence of 1 ms voltage write pulses of different amplitudes (from -V$_{MIN}$ to +V$_{MAX}$) while measuring the current during the application of these pulses. To record the HSL, we applied a small reading voltage (100 mV) after each writting pulse to determine the remanent resistance state. 

\begin{figure}[h!]
\centering
\includegraphics[scale=0.6]{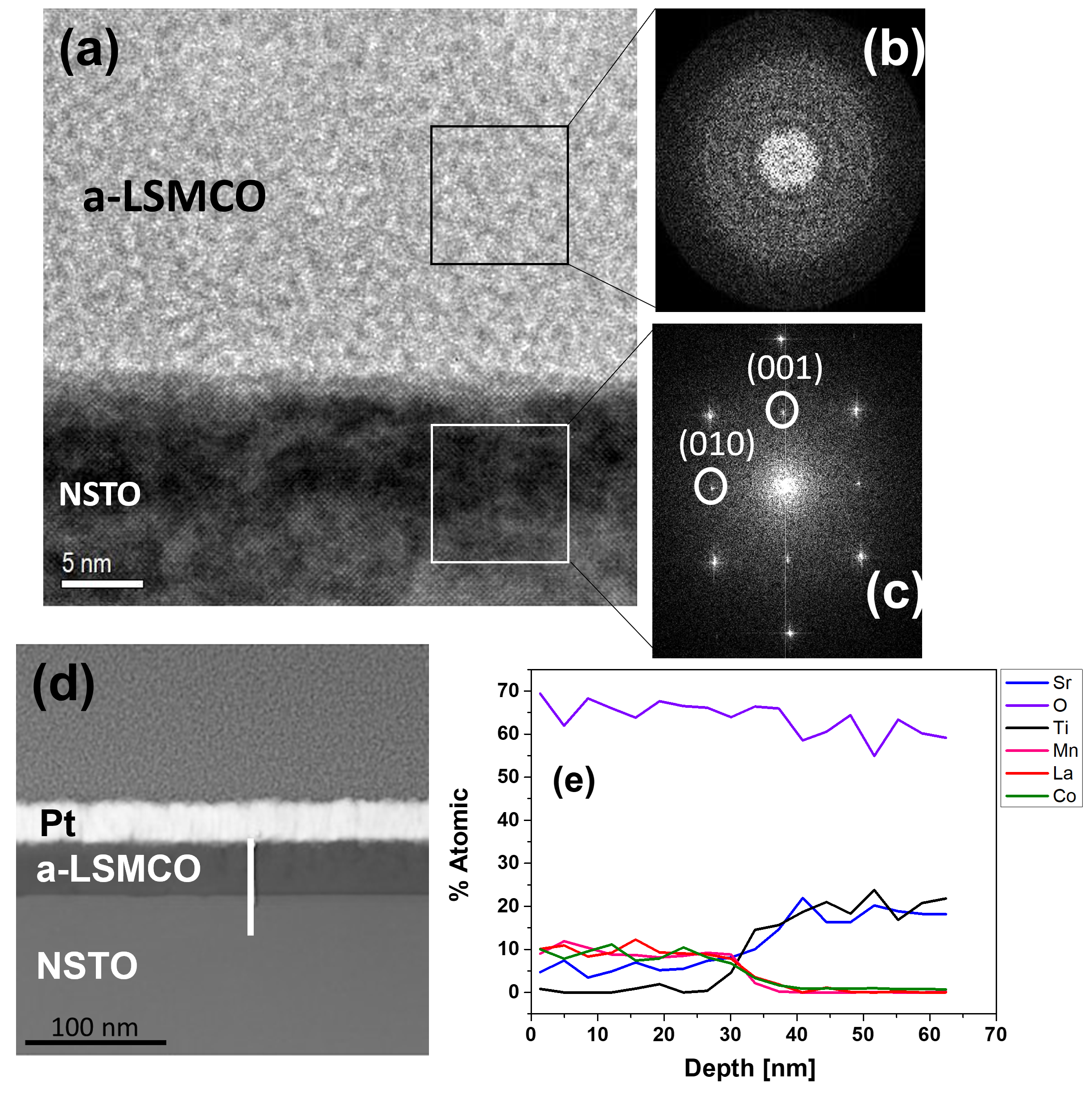}
\caption{(a) STEM-HAADF cross section corresponding to an amorphous LSMCO thin film on (001) NSTO; (b), (c) Fast Fourier transforms corresponding to selected areas of LSMCO and NSTO, displayed in panel (a), respectively. The amorphous nature of LSMCO is evidenced in the diffuse FFT without diffraction poles in comparison with the poles diffraction of the crystalline NSTO substrate. Zone axis is [100]; (d) STEM-HAADF cross section of a NSTO/amorphous LSMCO/Pt device; (e) EDS line scans for Sr, O, Ti, Mn, La and Co elements. The scan, indicated in (e) with a white line, starts at the LSMCO/Pt interface and ends in the NSTO substrate}
\label{fig2}
\end{figure}

\begin{figure*}[t]
\centering
\includegraphics[scale=0.5]{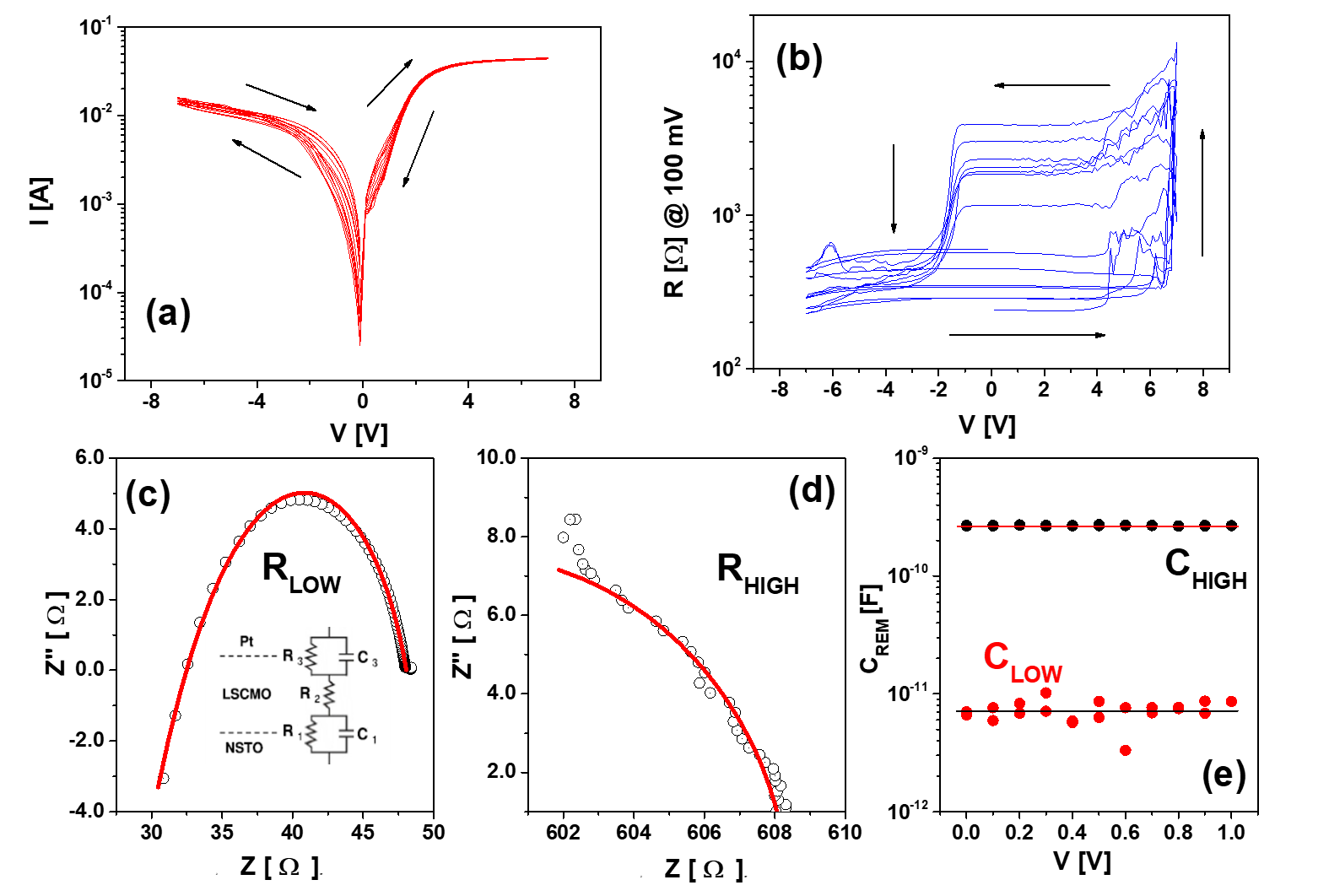}
\caption{(a) Dynamic I-V curves corresponding to an epitaxial (001) NSTO/LSMCO/Pt device stimulated with voltage. Ten consecutive cycles are shown; (b) Hysteresis switching loops corresponding to the same device and recorded simultaneously with the I-V curves shown in (a). The arrows indicate the evolution of the curves; (c) and (d) Impedance spectra measured on R$_{\mathrm{LOW}}$ and R$_{\mathrm{HIGH}}$ states, respectively. The inset in (c) shows the equivalent circuit used to fit the spectra. The fittings are shown in red lines; (e) Remanent  capacitance measured on the same device after the application of write pulses between 0 and 1 V. Two clear capacitive states are seen, C$_{\mathrm{HIGH}}$ corresponding to R$_{\mathrm{LOW}}$ and C$_{\mathrm{LOW}}$ corresponding to R$_{\mathrm{HIGH}}$.}
\label{fig3}
\end{figure*}

Figures 3(a) and (b) display, respectively, the dynamic I-V curve and HSL corresponding the LSMCO1 sample. Ten consecutive cycles are shown. It is seen that the dynamic I-V curves show rectifying behavior associated to the formation of a n-p diode at the NSTO-LSMCO interface (we recall that NSTO and LSMCO are n- and p-type materials, respectively \cite{tomio94,roman2020}). We also notice that, from the p-character of LSMCO and the high work function of Pt (5.6 eV), we expect a LSMCO/Pt quasi-ohmic interface. It is found that the device switches from low (R$_{\mathrm{LOW}}$ $\approx$ 400 $\Omega$) to high (R$_{\mathrm{HIGH}}$ $\approx$ 3 k$\Omega$) resistance (SET process) with positive stimuli, while the opposite transition (RESET process) is seen with negative stimuli. The HSL displayed in panel (b) shows that SET and RESET voltages are stable upon consecutive cycling, but both R$_{\mathrm{LOW}}$ and R$_{\mathrm{HIGH}}$ present some noticeable dispersion, an issue that will be addressed later. The device presents retention times for both resistive states of at least 10$^4$ s (not shown here). From the analysis and modelling of dynamic I-V curves, we showed that the memristive behavior originates at the NSTO/LSMCO interface upon oxidation/reduction of LSMCO \cite{roman2020}. We have successfully simulated the experimental HSL by assuming that LSMCO is in contact with an oxygen reservoir that allows the necessary oxygen release and uptake for the topotactic redox behavior, as shown in Ref. \cite{roman2020}. This scenario was further confirmed by electrically cycling the device in vacuum, where no SET process is observed, due to the impossibility of LSCMO to take atmospheric oxygen to become oxidized. 

\begin{figure}[t]
\centering
\includegraphics[scale=0.6]{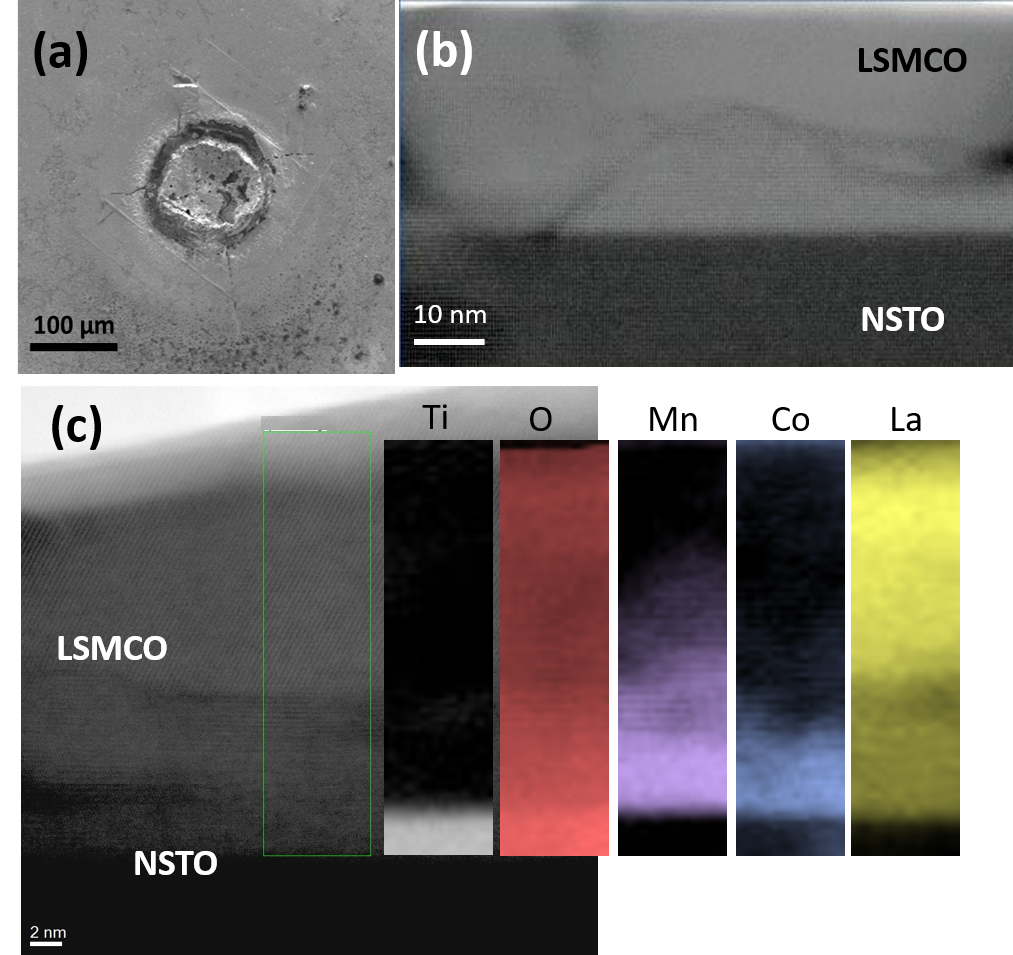}
\caption{(a) Scanning electron microscopy top-view for a formed epitaxial (001) NSTO/LSMCO/Pt device stimulated with voltage. A darker ring around the landing zone of the electrical tip, related to O$_2$ release and LSMCO/Pt expelling, is observed; (b) STEM-HAADF cross section corresponding to the central zone of the image shown in (a). The original epitaxial structure recrystallizes upon electroforming in an arrangement of non-coherent perovskite nanograins; (c) EELS maps corresponding to Ti, O, Mn, Co and La elements, for a formed (001) NSTO/LSMCO/Pt device. Changes in the brightness of the maps indicate chemical changes in the LSMCO layer upon forming (see text for details).}
\label{fig4}
\end{figure}

A very interesting property arises from the analysis of complex impedance spectra, shown in Figures 3(a) and 3(b) for R$_{\mathrm{LOW}}$ and R$_{\mathrm{HIGH}}$, respectively. The R$_{\mathrm{LOW}}$ spectrum was fitted by assuming the AC equivalent circuit shown in Figure 3(c) and the same circuit was used to fit the R$_{\mathrm{HIGH}}$ state. The values of the fitted circuit elements are shown in Table 1. It is found that the memristive behavior is mainly driven by changes in R$_1$, which changes between $\approx$ 10$\Omega$ and $\approx$ 500$\Omega$, and we associate to the resistance of the NSTO/LSMCO interface. The change in R$_1$ is concomitant with a large change of the interface capacitance C$_1$, which switches between $\approx$ 1.5 pF for R$_{\mathrm{LOW}}$ and $\approx$ 34 nF for R$_{\mathrm{HIGH}}$. The existence of a large memcapacitive effect is confirmed by measuring the remanent -overall- device capacitance C$_{REM}$ after the application of voltage write pulses between 0 and 1 V (small enough to avoid triggering a RESET transition). These measurements are shown in Figure 3(e) and display a C$_{\mathrm{HIGH}}$/C$_{\mathrm{LOW}}$ ratio of $\approx$ 100, that is an order of magnitude higher than the figures reported so far \cite{wu2011,sala2013,salarou2014,besso15,Yang_2017,park2018,Liu_2018, boro2020,guo2020}. We notice that C$_{REM}$ is a function of the equivalent circuit elements (Table 1) and the frequency $\omega$, as described by the Maxwell-Wagner model \cite{catalan2006}. 
We relate the observed multi-mem behavior to the oxidation/reduction of LSMCO. An oxidized interface with NSTO results in a R$_{\mathrm{LOW}}$-C$_{\mathrm{HIGH}}$ state, while a reduced interface leads into a R$_{\mathrm{HIGH}}$-C$_{\mathrm{LOW}}$ state. The large difference in oxygen content between both states critically affects the balance between n and p carriers at the NSTO/LSCMO interface, which controls the n-p junction depletion layer and triggers the multi-mem behavior.
\begin{table}[h!]
\caption {Numeric values of the elements of the equivalent circuits used to fit the impedance spectra for all samples. Resistances are given either on $\Omega$ or k$\Omega$ and capacitances on pF or nF.  }
\label{tab} 
\begin{tabular}{|c|c|c|c|c|c|c|}
\hline
\multicolumn{1}{|c|}{\textbf{Sample}}        & \multicolumn{1}{|c|}{\textbf{State}} & \multicolumn{1}{|c|}{\textbf{R$_1$ [$\Omega$]}} & \multicolumn{1}{|c|}{\textbf{C$_1$[F]}} & \multicolumn{1}{|c|}{\textbf{R$_2$ [$\Omega$]}} & \multicolumn{1}{|c|}{\textbf{R$_3$ [$\Omega$]}} & \multicolumn{1}{|c|}{\textbf{C$_3$[F]}}                                                    \\ 
\hline
LSMCO1 &R$_{\mathrm{HIGH}}$& 500                                & 1.5p                          & 90 & 18 & 6n                                                                                \\
\hline
 &R$_{\mathrm{LOW}}$& 10                                & 34n                           & 17 & 21 & 3n                                                                                 \\

\hline
LSMCO2 &R$_{\mathrm{HIGH}}$& 79.9k                                & 0.12n                           & 70 & -- & --                                                                         \\ 
\hline
 &R$_{\mathrm{LOW}}$& 27.7k                                & 0.14n                           & 70 & -- & --                                                                   \\ 
\hline
LSMCO3 &R$_{\mathrm{HIGH}}$& 8.6k                                & 8p                          & 55 & 340 & 2.3n                                                                          \\ 
\hline
  &R$_{\mathrm{LOW}}$& 70                                & 0.55n                          & 15 & 22 & 2.5n      
                                                        \\ 
\hline
LSMCO4 &R$_{\mathrm{HIGH}}$& 1.3k                                & 3p                          & 213 & 60 & 6.6n                                                                           \\ 
\hline
  &R$_{\mathrm{LOW}}$& 16                                & 49n                          & 53 & 23 & 4.1n                       
                                                        \\ 
\hline
\end{tabular}
\end{table}

In Figure 4 we analyze the structural and chemical changes produced on the device upon the application of electrical stress. Figure 4(a) shows a scanning electron microscopy top view of a formed device, where it is evidenced the presence of a ring (darker contrast) around the landing zone of the tip used to make electrical contact. This ring is a signature of oxygen release upon LSMCO reduction. Also, LSMCO and Pt are expelled upon O$_2$ release, electrically decoupling the central zone from the rest of the device. Figure 4(b) shows a STEM-HAADF cross section of the formed device, located in the central zone of Figure 4(a). It is seen that the initial epitaxial LSMCO structure changes to an arrange of non-coherent perovskite nanograins.  We have previously shown that the grains in contact with the NSTO substrate correspond to oxidized LSMCO, but some of the grains in contact with the top Pt electrode present oxygen deficiency \cite{roman2020}. Also, other chemical changes are observed for formed LSCMO, as it is displayed in the EELS maps of Figure 4(c). It is found that the grains in contact with the NSTO substrate retain the LSMCO cation stoichiometry but the grains in contact with the top Pt electrode present a deficiency of Mn and Co and an excess of La. This structural and chemical changes are produced by uncontrolled power release and Joule heating upon the application of voltage pulses. Later we will discuss different strategies to minimize these effects.

\begin{figure*}[t]
\centering
\includegraphics[scale=0.6]{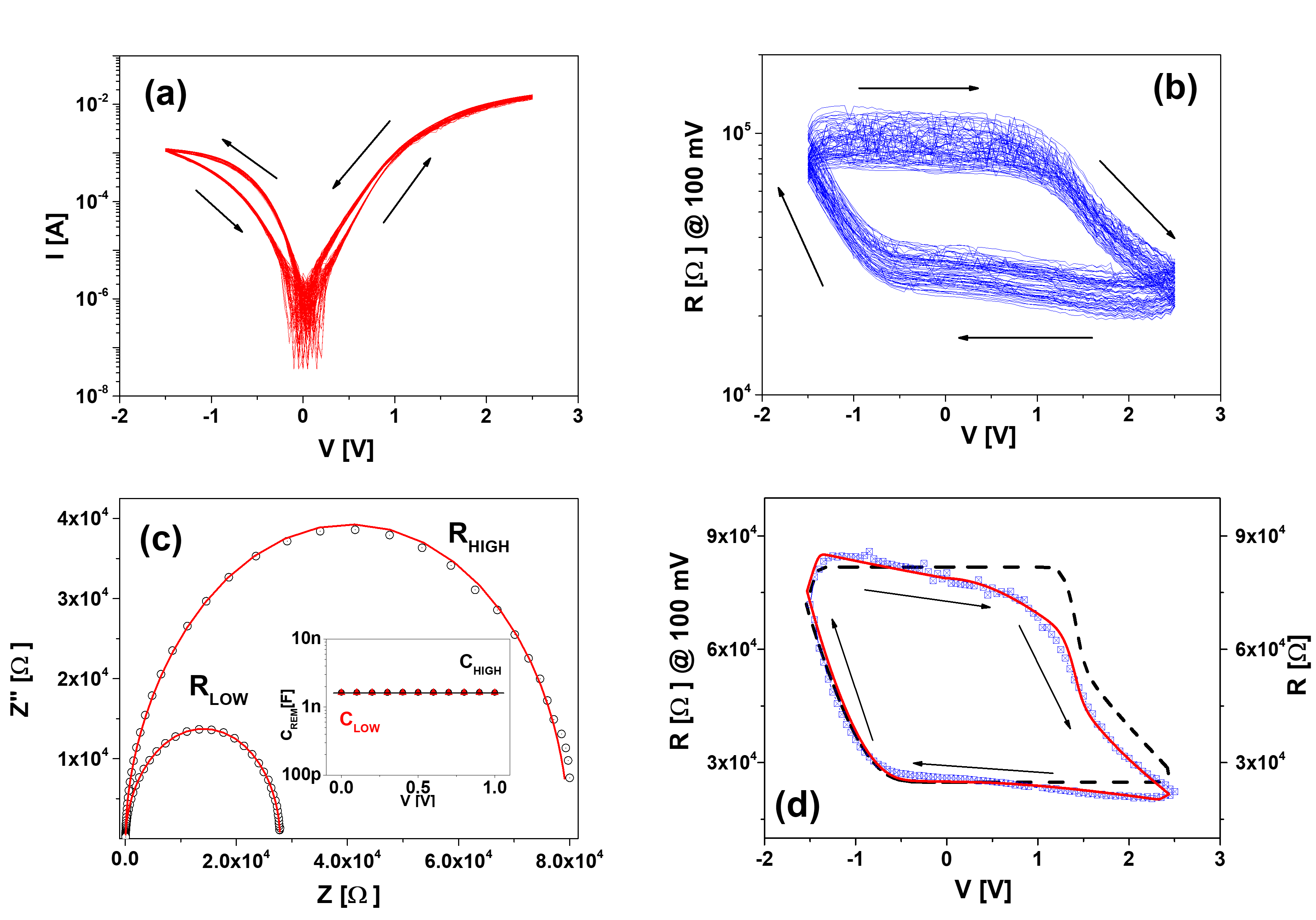}
\caption{a) Dynamic I-V curves corresponding to an amorphous NSTO/LSMCO/Pt device stimulated with voltage. 100 consecutive cycles are shown; (b) Hysteresis switching loops corresponding to the same device and recorded simultaneously with the I-V curves shown in (a). The arrows indicate the evolution of the curves; (c) Impedance spectra measured on R$_{\mathrm{LOW}}$ and R$_{\mathrm{HIGH}}$ states, respectively. The fitting of the spectra are shown as red lines. The inset shows the remanent  capacitance measured on the same device after the application of writing pulses between 0 and 1 V. No memcapacitance is found; (d) Simulated Hysteresis Switching Loop with (solid line) and without (dashed line) the inclusion of the non-linear circuit element, emulating the Poole-Frenkel conduction mechanism. The experimental HSL is shown with blue squared symbols}
\label{fig5}
\end{figure*}

We focus now on the electrical response of NSTO/amorphous LSMCO/Pt (LSCMO thickness was 36.5 nm). We will refer to this sample as LSMCO2. The sample virgin resistance was $\approx$ 100 M$\Omega$ and for the forming process we applied +4V pulses that lowered the device resistance to $\approx$ 20 k$\Omega$. Figures 5(a) and (b) display the dynamic I-V curves and HSL corresponding to LSMCO2 sample. The time-width of the writing pulses was around 1 ms. 100 consecutive cycles are shown with a stable behavior. Although rectifying behavior is also observed in LSMCO2 (see Figure 5(a)), the SET (RESET) transitions take place upon the application of a positive (negative) voltage (Figure 5(b)), contrary to the LSMCO1 sample, which displayed mirrored SET/RESET polarities (see Figure 3). This suggests a different memristive mechanism for the LSMCO2 device. 
R$_{\mathrm{{HIGH}}}$ and R$_{\mathrm{{LOW}}}$ states are $\approx$ 90 k$\Omega$ and $\approx$ 30 k$\Omega$, respectively, with a good reproducibility upon consecutive cycling. Retention times for both states were checked up to 10$^3$ s (not shown here). A completely different behavior with respect to LSMCO1 sample arises from the impedance spectra, which are shown for LSMCO2 sample in panel (c) of Figure 5. In this case, both spectra could be fitted by assuming a parallel resistor-capacitor combination (R$_1$, C$_1$) in series with another resistor (R$_2$). The numerical values of the fitted circuit elements are shown in Table 1. The striking difference with the case of LSCMO1 is the absence of memcapacitance, reflected in an unchanged value of C$_1$ $\approx$ 0.1 nF between R$_\mathrm{{HIGH}}$ and R$_{\mathrm{LOW}}$ states. This is confirmed by the evolution of the remanent capacitance C$_{REM}$ vs. writing voltage (range 0-1 V), displayed in the inset of Figure 5(c), where no changes are found between both resistive states (C$_{REM}$ $\approx$ 2 nF). 
In order to shed light on the microscopic origin of the distinct electrical behavior of LSMCO2 sample, we performed the experiments shown in  Figure 6. Figure 6(a) displays a scanning electron microscopy top-view of the device after forming. Although it is seen that the area of the top electrode in contact with the electrical tip used to apply the voltage was damaged, no indication of strong O$_2$ release is observed (compare with the case of the LSMCO1 sample, Figure 4(a)), suggesting the absence of LSMCO2 topotactic redox transition. Further information can be obtained from panels (b)-(d), displaying a high resolution STEM-HAADF cross-section of the formed device (b) and EDS line scans from the film's interface with Pt to the NSTO substrate ((c) and (d)). Several features are observed: i) a layer of about 12nm in thickness of the oxide in contact with the NSTO substrate becomes crystallized after forming; ii) the layer of about 24nm in thickness on top of the crystallized oxide layer remains amorphous and displays a different STEM-HAADF Z-contrast, suggesting that it presents a different chemistry in relation to the bottom layer. The latter is confirmed by the EDS line scans displayed in Figure 6(c) and (d), which indicate the presence of an oxide bilayer, consisting of a top layer of SrTiO$_3$ (STO) and a bottom layer with a stoichiometry close to the double-perovskite La$_2$CoMnO$_6$ (LCMO). The presence of LCMO -which does not present a topotactic transition and was reported to increase its conductivity with the OV content \cite{sayed2014}- at the interface with NSTO explains the absence of memcapacitance for the LSMCO2 device.  

Regarding the memristive behavior, we have successfully reproduced  in Figure \ref{fig5}(d) the experimental remanent resistance loop  by modelling  the OV dynamics between STO and LCMO layers using the Voltage-Enhanced-Oxygen Vacancies (VEOV) model \cite{rozenberg_2010,ghenzi_2013}.
Both STO and LCMO behave as n-type semiconductors in which OV act as electronic dopants reducing  their resistivities \cite{sayed2014,spinelli2010}. We assume the LCMO/NSTO interface as ohmic and  therefore not contributing to the memristive effect. On the other hand, the Pt/STO interface is of Schottky-type \cite{mikheev2015} and  favours the generation of strong electric fields upon the application of electrical stress, promoting OV electromigration between STO and LCMO. The simulation assumes a 1D chain of LCMO  and STO  nanodomains, able to accommodate different OV contents. Each nanodomain is caracterized by a resistivity which is related to the local OV density. For a given value of the applied electrical stimulus, in each simulation step the OV profile is updated through a set of balance equations for the OV  transition rates  and the resistance of the sample is computed \cite{rozenberg_2010}. In the present case the chain of nanodomains is divided into three regions that define two interfaces: the Schottky  Pt/STO interface (LI) and the STO/LCMO  interface (RI), respectively. Upon the application of positive electrical stimuli to the Pt top electrode, OV electromigrate from the LI to the RI, with the concomitant reduction of the total resistance.   
For a negative polarity of the applied stress, the opposite process takes places and the resistance increases.  The simulations produce a squared remanent resistance loop, shown by the dashed  line in  Figure 5(d), with the same circulation than the experimental HSL. However  to fully capture the slopes  of both R$_{\mathrm{LOW}}$ and R$_{\mathrm{HIGH}}$ states exhibited in Figure 5(b) an additional (non-linear) circuital  element has been included  in parallel with the memristive channel. We speculate that an electronic Poole-Frenkel (PF) conduction mechanism, possibly linked to the STO layer, coexists with the memristive channel, ruled  by the OV dynamics. We recall that PF is a bulk conduction mechanism related to the electronic emission of carriers from traps in the oxide. In the present case, these traps may be present in the STO amorphous layer. Figure 5(d) also displays the simulation of the remanent resistance loop (solid line) after the inclusion of the non-linear element emulating the PF channel. The agreement with the experimental curve is remarkable.

\begin{figure}[t]
\centering
\includegraphics[scale=0.6]{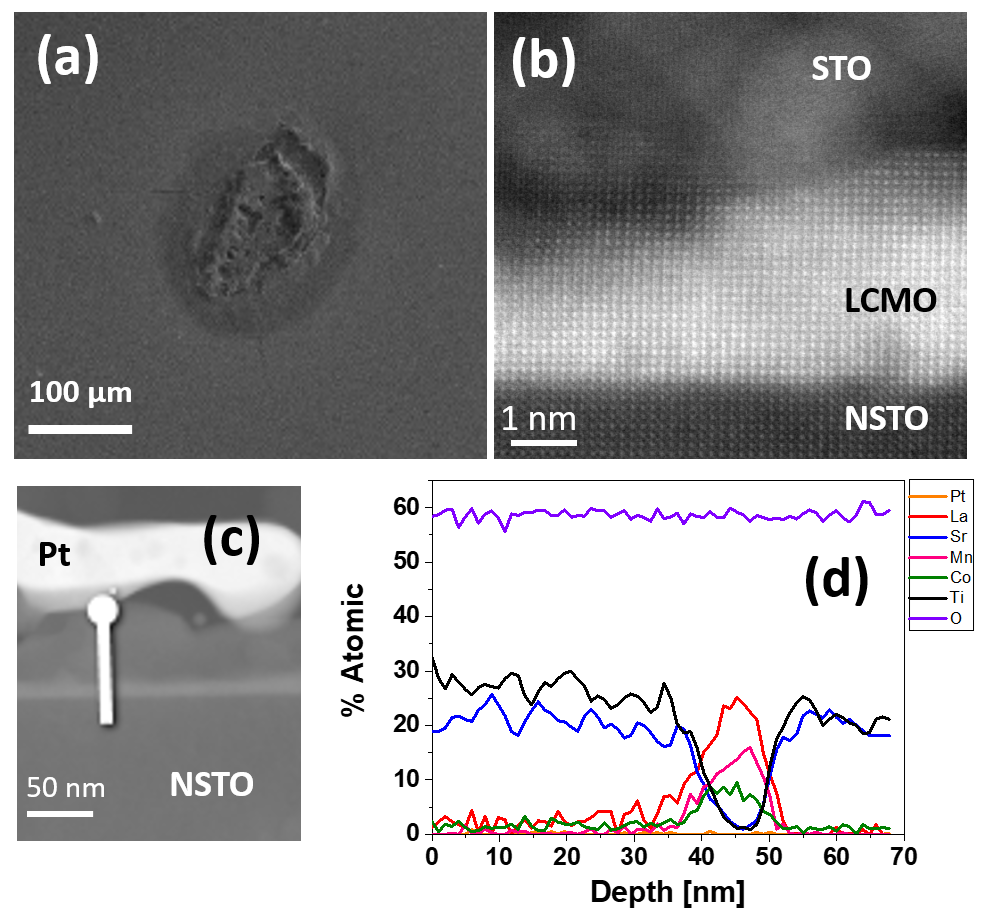}
\caption{(a) Scanning electron microscopy top-view of a formed amorphous NSTO/LSMCO/Pt device stimulated with voltage. No evidence of strong O$_2$ release is found; (b) High resolution STEM-HAADF cross section corresponding to the same device. The lamella was prepared close to the border of the fused Pt zone shown in (a); (c), (d) EDS line scans corresponding to Pt, La, Sr, Mn, Co, O and Ti elements, for the same device. As indicated in (c), the scan starts at the interface between the film and the Pt top electrode and ends in the NSTO substrate. The formation of a bottom crystalline LCMO layer and an amorphous STO one is observed (see text for details).}
\label{fig6}
\end{figure}

In summary, in this Section we have shown, for both epitaxial and amorphous LSMCO thin films, that the application of voltage stress produces controlled mem-behavior but this is accompanied with a strong structural damage and  changes in the nanostructure and chemistry of the active layers. In the next Section we analyze different strategies to circumvent these unwanted effects.

\section{CHARACTERIZATION OF CURRENT-CONTROLLED DEVICES}
 
In this section, we address different strategies to maintain the multi-mem behavior observed in epitaxial NSTO/LSMCO interfaces stimulated with voltage (sample LSMCO1) minimizing at the same time the nanostructural and chemical changes observed upon the application of electrical stress. First, we repeated the set of electrical experiments already described for LSMCO1 using \emph{current} pulses instead of voltage ones. In the case of the application of voltage writing pulses, both the forming and SET transition imply a decrease of the device resistance together with an overshoot of injected power while the applied voltage is maintained (given the well-known relationship P= V$^2$/R), which is likely responsible for the self-heating effects that drive the structural damage and chemical diffusion. If current is used as stimuli during the forming and SET transition, the injected electrical power (P=I$^2$R) remains self-limited and, as we have shown for other manganite-related memristive devices \cite{acevedo2016,acevedo2017}, a subtler control of the device resistive changes is gained. We will refer to the epitaxial (001) NSTO/LSMCO/Pt device controlled with current as LSMCO3, presenting a LSMCO thickness of 13 nm. The forming process of the device was achieved by applying a sequence of current pulses of -200 mA which changed the virgin resistance from $\approx$ 100 k$\Omega$ to $\approx$ 50$ \Omega$. Figures 7(a) and (b) display the dynamic I-V curve and remanent resistance loop corresponding to this sample. The time-width of the writing pulses was around 1ms. As in the case of LSMCO1 sample, we observed from the I-V curve a rectifying behavior, related to the NSTO/LSMCO n-p interface and a memristive effect characterized by RESET (SET) transitions after the application of positive (negative) current pulses. R$_{\mathrm{HIGH}}$ and R$_{\mathrm{LOW}}$ states are $\approx$ 1 k$\Omega$ and $\approx$ 300 $\Omega$, respectively. Upon consecutive cycling, the stability and repeatability of the memristive effect are significantly enhanced in relation to LSMCO1 sample, as shown in Figures 7(a) and (b) for 50 cycles. We will come later to this issue. Retention times for both states were tested up to 10$^3$ s (not shown here). Figures 7(c) and (d) display impedance spectra corresponding to R$_{\mathrm{LOW}}$ and R$_{\mathrm{HIGH}}$ states, respectively. The spectra were fitted by assuming the same equivalent circuit as in the case of LSMCO1 device, and the extracted values for the circuit elements are shown in Table 1. As in the case of LSMCO1 device, we see the presence of a strong memcapacitive behavior reflected in a change of C$_1$ between $\approx$ 8.5 pF for R$_{\mathrm{HIGH}}$ and $\approx$ 0.5 nF for R$_{\mathrm{LOW}}$. Concomitantly, R$_1$ changes between $\approx$ 8.5 k$\Omega$ and $\approx$ 70 $\Omega$, respectively. We attribute the changes in R$_1$ and C$_1$ to the NSTO/LSMCO interface. The memcapacitive effect is confirmed by Figure 7(e), which shows remanent capacitance C$_{REM}$ - voltage measurements for both capacitive states, where a C$_{\mathrm{HIGH}}$/C$_{\mathrm{LOW}}$ ratio of $\approx$ 100 is observed. These measurements demonstrate that the multi-mem behavior between LSMCO1 and LSMCO3 devices is similar, and likely arises from the same mechanism. 
\begin{figure*}[t]
\centering
\includegraphics[scale=0.6]{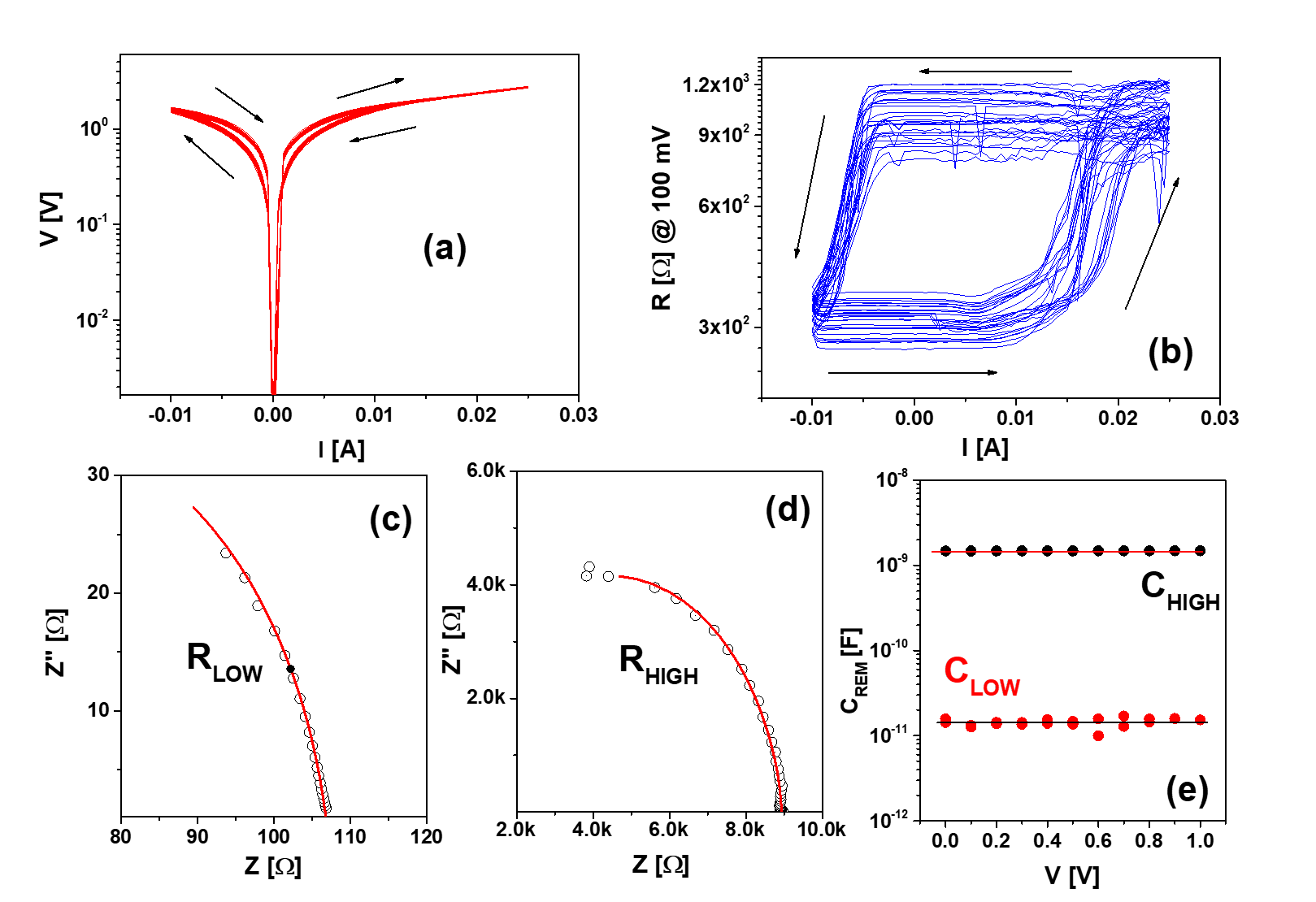}
\caption{(a) Dynamic I-V curves corresponding to an epitaxial (001) NSTO/LSMCO/Pt device stimulated with current pulses. 50 consecutive cycles are shown; (b) Hysteresis switching loops corresponding to the same device and recorded simultaneously with the I-V curves shown in (a). The arrows indicate the evolution of the curves; (c) and (d) Impedance spectra measured on R$_{\mathrm{LOW}}$ and R$_{\mathrm{HIGH}}$ states, respectively; (e) Remanent capacitance-voltage curve measured on the same device for the two capacitive states. We notice that C$_{\mathrm{HIGH}}$ corresponds to R$_{\mathrm{LOW}}$ and C$_{\mathrm{LOW}}$corresponds to R$_{HIGH}$.}
\label{fig7}
\end{figure*}

\begin{figure}[t]
\centering
\includegraphics[scale=0.6]{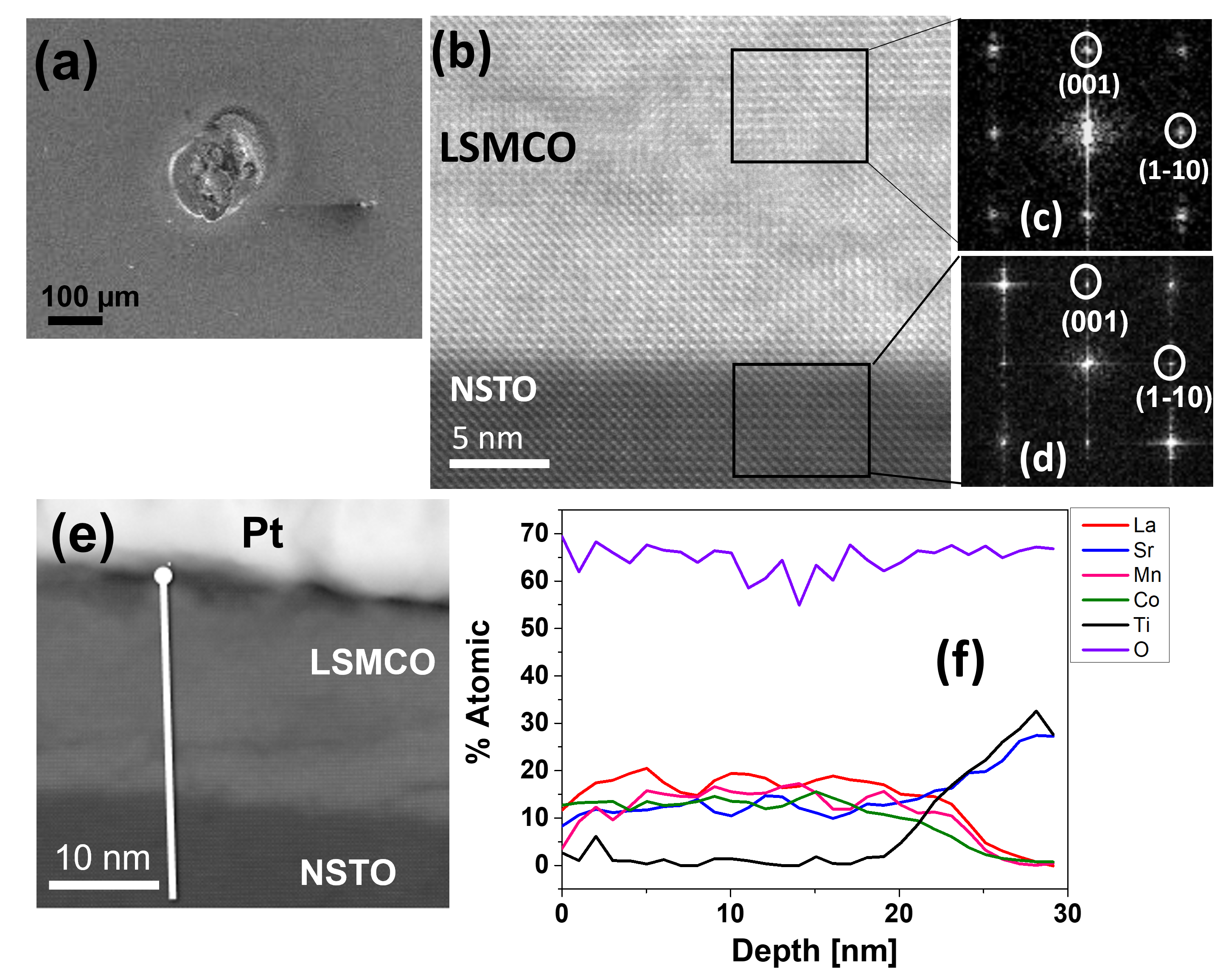}
\caption{(a) Scanning electron microscopy top-view of a formed epitaxial (001) NSTO/LSMCO/Pt device stimulated with current. No evidence of strong O$_2$ release is found; (b) STEM-HAADF cross section corresponding to the same device. The lamella was prepared close to the border of the fused Pt zone shown in (a); (c), (d) FFTs corresponding to selected areas of the LSMCO layer and the NSTO substrate, respectively. Zone axis is [110]. The structural coherence between NSTO and LSMCO is maintained upon forming; (e) Lower magnification STEM-HAADF image corresponding to the same device. The presence of some extended defects (darker contrast) in the LSMCO layer is seen; (f) EDS line scans corresponding to La, Sr, Mn, Co, O and Ti elements, for the formed device. As panel (e) shows, the scan starts at the LSMCO/Pt interface and end in the NSTO substrate. The cation stochiometry of the virgin device is maintained, within the error of the technique.}
\label{fig8}
\end{figure}

\begin{figure*}[t]
\centering
\includegraphics[scale=0.6]{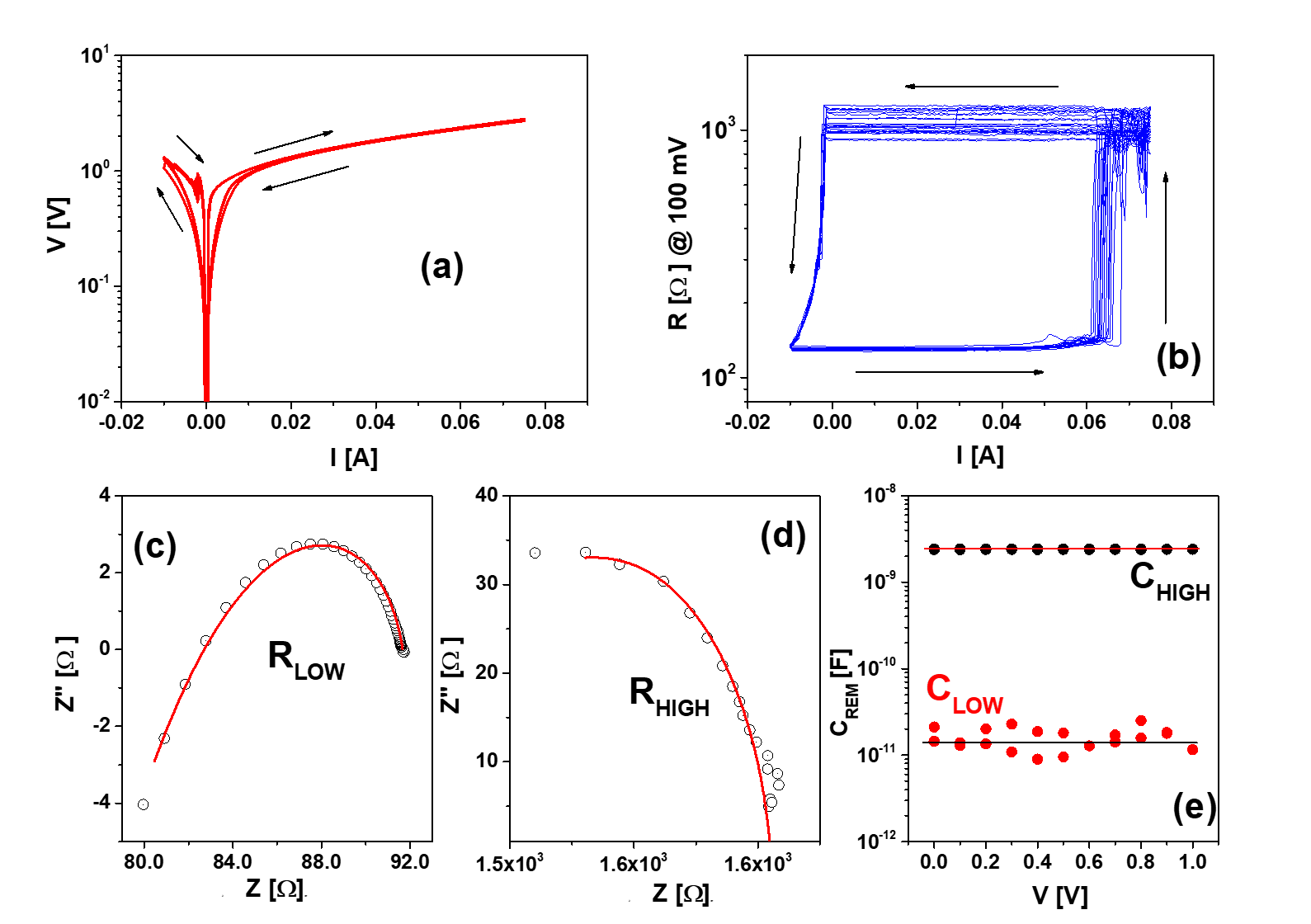}
\caption{(a) Dynamic I-V curves corresponding to an epitaxial (110) NSTO/LSMCO/Pt device stimulated with current pulses. 50 consecutive cycles are shown; (b) Hysteresis switching loops corresponding to the same device and recorded simultaneously with the I-V curves shown in (a). The arrows indicate the evolution of the curves; (c) and (d) Impedance spectra measured on R$_{\mathrm{LOW}}$ and R$_{\mathrm{HIGH}}$ states, respectively; (e) Remanent capacitance-voltage curve measured on the same device for the two capacitive states. We notice that C$_{\mathrm{HIGH}}$ corresponds to R$_{\mathrm{LOW}}$ and C$_{\mathrm{LOW}}$ corresponds to R$_{HIGH}$.}
\label{fig9}
\end{figure*}

\begin{figure*}[t]
\centering
\includegraphics[scale=0.7]{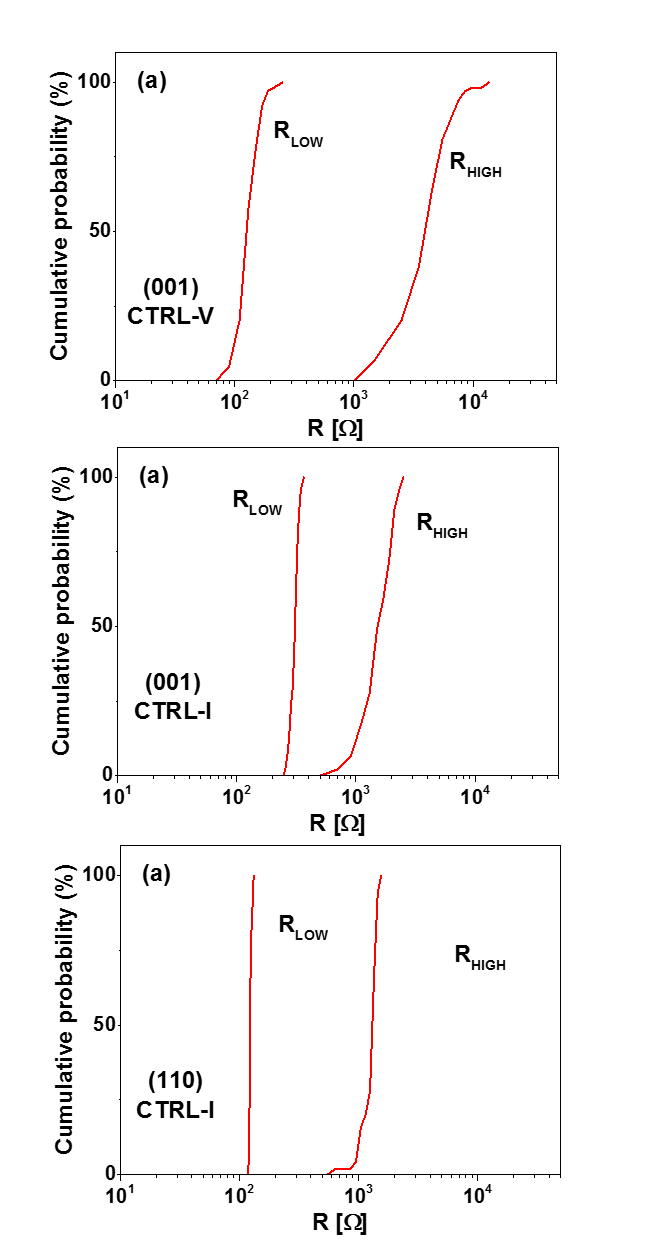}
\caption{Cumulative probabilities related to consecutive cycling -200 cycles- between R$_{\mathrm{LOW}}$ and R$_{\mathrm{HIGH}}$ for: (a) (001) NSTO/LSMCO/Pt device stimulated with voltage, (b) (001) NSTO/LSMCO/Pt device stimulated with current and (c) (110) NSTO/LSMCO/Pt device stimulated with current.}
\label{fig10}
\end{figure*}

In order to observe the structure and chemistry of the formed LSMCO3 device at the nanoscale, we have performed the experiments shown in Figure 8. Figure 8(a) displays a scanning electron microscopy top-view of a formed device. It is found that the zone of contact with the electrical tip is melted but, unlike the case of the voltage-controlled LSMCO1 device, no material expelling caused by oxygen release is seen. This is confirmed by the STEM-HAADF cross-section of Figure 8(b) and the corresponding FFTs displayed in panels (c) and (d), which show, despite the formation of some extended defects (see also panel (e)), that the structural coherence between the NSTO substrate and the LSMCO layer is conserved. Moreover, the EDS line scans displayed in Figures 8(e) and (f) indicate that the cation stoichiometry of the pristine LSMCO layer is conserved upon forming, unlike in the case of the LSMCO1 device, where strong cation diffusion was observed. These experiments show that the multi-mem behavior observed in LSMCO1 was maintained in LSMCO3 upon electrical cycling, but in the latter the structural damage linked to oxygen release and the nanostructural and chemical changes upon electrical stressing are minimized by the strategy of controlling the power dissipation by using current pulses.  

Further improvement in the electrical response of the devices can be achieved by using (110) epitaxial NSTO/LSMCO/Pt structures. We will describe the properties of a device with a LSMCO thickness of 16.3 nm, labelled as LSMCO4. Stimuli with current pulses (time-width of 1 ms) were maintained. The resistance of the virgin device was $\approx$ 30 k$\Omega$, which switches to $\approx$ 80 $\Omega$ after the application of -220 mA pulses . The dynamic I-V curve of Figure 9(a) and the HSL of Figure 9(b), recorded for 50 cycles, are similar to those corresponding to the LSMCO3 device. R$_{\mathrm{HIGH}}$ and R$_{\mathrm{LOW}}$ states are $\approx$ 1 k$\Omega$ and $\approx$ 150 $\Omega$, respectively; however, a more robust and stable memristive behavior that the one found for the LSMCO3 device is obtained. Retention times for both resistive states were checked up to 10$^3$ s (not shown here). The memcapacitive behavior is also conserved in the LSMCO4 device, as inferred from the impedance spectra displayed in Figs. 9(c) and (d). The same equivalent circuit than those used for the LSMCO1 and LSMCO3 devices was used to fit the spectra, and the extracted values for the circuit elements are shown in Table 1. Again, a multi-mem behavior linked to the NSTO/LSMCO interface occurs, reflected in a change of the interface capacitance C$_1$ between $\approx$ 3 pF for R$_{\mathrm{HIGH}}$ and $\approx$ 50 nF for R$_{\mathrm{LOW}}$. Concomitantly, the interface leakage channel, represented by R$_1$, changes between $\approx$ 1.3 k$\Omega$ and $\approx$ 15 $\Omega$, respectively. The  device remanent capacitances display a ratio C$_{\mathrm{HIGH}}$/C$_{\mathrm{LOW}} \approx$ 140, shown in Figure 9(d), about 40 $\%$ larger than in the case of the LSCMO3 device. Microscopy analysis (not shown here) evidenced that upon forming there is no significant structural damage related to oxygen release, the epitaxial character of the active LSMCO layer is maintained and no significant cation diffusion was detected, as was also the case of the LSMCO3 device. 

The reason for the improved electrical response of LSCMO4 device in relation to LSMCO3 is likely related to the fact that the (110) out-of-plane orientation of the former allows crystallographic planes parallel to the [001] brownmillerite axis \cite{nota} to bridge the NSTO substrate with the Pt top electrode \cite{acharya2017,mou2021,kim2020,nalla2019,nalla2020}. These planes are known to behave as fast paths for oxygen drift and, therefore, ease the uptake and release of oxygen from/to the environment upon electrical cycling. 

Based on the electrical and microscopy results previously presented for epitaxial devices, we see a progressive improvement of the electrical response from LCMO1 to LCMO2 and then to LCMO3. In order to provide further evidence of this, we performed endurance experiments on the three devices, consisting in switching back and forth between R$_{\mathrm{HIGH}}$ and R$_{\mathrm{LOW}}$ states with single pulses during 200 cycles. The voltage/current of the pulses were chosen in each case to obtain a stable switching. For the LSMCO1 device, single SET and RESET pulses of -4 V and +7 V were applied, respectively. For LSMCO3, the SET and RESET pulses were of -10 mA and +30 mA, while for LSMCO4 they were of -10 mA and +70 mA, respectively. From the endurance curves (not shown here) we extracted the cumulative probabilities linked to both R$_{\mathrm{HIGH}}$ and R$_{\mathrm{LOW}}$ states, as shown in Figure 10 for the three devices. It is evident that the cumulative probabilities become narrower when going from LSMCO1 to LSMCO2 and then to LSMCO3 devices, evidencing a more stable and reproducible mem-behavior. This suggests that both strategies -using current pulses as write stimuli and an (110) out-of-plane orientation- favour the mem-effect, strongly limiting the nanostructural and chemical modifications of the device active LSMCO layer upon the application of electrical stress.

\section{CONCLUSIONS}

In summary, we have performed a systematic study of the electrical behavior of multi-mem NSTO/LSMCO/Pt devices with different crystallinity, out-of-plane orientation and stimulation (either with voltage or current pulses). The voltage-driven epitaxial and amorphous LSMCO devices grown on (001) NSTO -labelled as LSMCO1 and LSMCO2- show completely different mem-mechanisms. In the first case, we found a memristive and memcapacitive response, linked to the topotactic oxidation and reduction of the LSMCO layer in contact with NSTO, which is know to form a switchable n-p diode \cite{roman2020}. In the second one, chemical modifications upon forming leave a double perovskite (non-topotactic) LCMO epitaxial layer in contact with NSTO and an amorphous STO layer in contact with the Pt top electrode. The absence of LSMCO in contact with NSTO is likely the reason for the absence of memcapacitance, giving strong support to the multi-mem mechanism proposed for the LSMCO1 device. The memristive effect suggested for the LSMCO2 device is related to a more standard mechanism of OV exchange between LCMO and STO layers. This has been confirmed by modelling the OV dynamics between both layers. The comparison between epitaxial (001) NSTO/LSMCO/Pt devices stimulated with voltage (LSMCO1) or current (LSMCO3) shows a more robust and stable mem-response for the latter. We attribute this improvement to the self-limited electrical power injection upon forming in the case of current-controlled devices, which avoids the structural damage linked to oxygen release seen on LSMCO1 device and minimizes nanostructural and chemical changes, all linked to self heating effects. Further improvement in the electrical response can be obtained for (110) NSTO/LSMCO/Pt devices (LCMO4), which we attribute to easy oxygen migration in and out of the device -necessary to achieve the topotactic redox transitions- through planes parallel to the [001] brownmillerite axis, which bridge the NSTO substrate with the Pt top electrode, similar to what it has been reported in other systems \cite{acharya2017,mou2021,kim2020,nalla2019,nalla2020}. Our work helps paving the way for the integration of LSMCO-based devices in cross-bar arrays, in order to exploit their memristive and memcapacitive properties for the development of neuromorphic or in-memory computing devices.

\section*{Acknowlegments}
 
We acknowledge support from UNCuyo (06/C591), ANPCyT (PICT2017-1836, PICT2019-02781 and PICT2019-0654) and EU-H2020-RISE project "MELON" (SEP-2106565560). This publication is part of the project "Memcapacitive elements for cognitive devices", project number 040.11.735, which is financed by the Dutch Research Council (NWO). We acknowledge support from the Ministry of Science and Higher Education of the RF (state contract N13.2251.21.0042). We also acknowledge the LMA-Universidad de Zaragoza for offering access to the microscopy instruments.     

\section*{DATA AVAILABILITY}

The data that support the findings of this study are available from the corresponding author
upon reasonable request.

\appendix

 

\bibliography{references1}

\begin{thebibliography}{43}%
\makeatletter
\providecommand \@ifxundefined [1]{%
 \@ifx{#1\undefined}
}%
\providecommand \@ifnum [1]{%
 \ifnum #1\expandafter \@firstoftwo
 \else \expandafter \@secondoftwo
 \fi
}%
\providecommand \@ifx [1]{%
 \ifx #1\expandafter \@firstoftwo
 \else \expandafter \@secondoftwo
 \fi
}%
\providecommand \natexlab [1]{#1}%
\providecommand \enquote  [1]{``#1''}%
\providecommand \bibnamefont  [1]{#1}%
\providecommand \bibfnamefont [1]{#1}%
\providecommand \citenamefont [1]{#1}%
\providecommand \href@noop [0]{\@secondoftwo}%
\providecommand \href [0]{\begingroup \@sanitize@url \@href}%
\providecommand \@href[1]{\@@startlink{#1}\@@href}%
\providecommand \@@href[1]{\endgroup#1\@@endlink}%
\providecommand \@sanitize@url [0]{\catcode `\\12\catcode `\$12\catcode
  `\&12\catcode `\#12\catcode `\^12\catcode `\_12\catcode `\%12\relax}%
\providecommand \@@startlink[1]{}%
\providecommand \@@endlink[0]{}%
\providecommand \url  [0]{\begingroup\@sanitize@url \@url }%
\providecommand \@url [1]{\endgroup\@href {#1}{\urlprefix }}%
\providecommand \urlprefix  [0]{URL }%
\providecommand \Eprint [0]{\href }%
\providecommand \doibase [0]{http://dx.doi.org/}%
\providecommand \selectlanguage [0]{\@gobble}%
\providecommand \bibinfo  [0]{\@secondoftwo}%
\providecommand \bibfield  [0]{\@secondoftwo}%
\providecommand \translation [1]{[#1]}%
\providecommand \BibitemOpen [0]{}%
\providecommand \bibitemStop [0]{}%
\providecommand \bibitemNoStop [0]{.\EOS\space}%
\providecommand \EOS [0]{\spacefactor3000\relax}%
\providecommand \BibitemShut  [1]{\csname bibitem#1\endcsname}%
\let\auto@bib@innerbib\@empty
\bibitem [{\citenamefont {Sawa}(2008)}]{saw_2008}%
  \BibitemOpen
  \bibfield  {author} {\bibinfo {author} {\bibfnamefont {A.}~\bibnamefont
  {Sawa}},\ }\href@noop {} {\bibfield  {journal} {\bibinfo  {journal}
  {Materials Today}\ }\textbf {\bibinfo {volume} {11}},\ \bibinfo {pages} {28}
  (\bibinfo {year} {2008})}\BibitemShut {NoStop}%
\bibitem [{\citenamefont {Ielmini}\ and\ \citenamefont
  {Waser}(2016)}]{iel_2016}%
  \BibitemOpen
  \bibfield  {author} {\bibinfo {author} {\bibfnamefont {D.}~\bibnamefont
  {Ielmini}}\ and\ \bibinfo {author} {\bibfnamefont {R.}~\bibnamefont
  {Waser}},\ }\href@noop {} {\emph {\bibinfo {title} {Resistive Switching: From
  Fundamentals of Nanoionic Redox Processes to Memristive Device
  Applications}}}\ (\bibinfo  {publisher} {Wiley-VCH},\ \bibinfo {year}
  {2016})\BibitemShut {NoStop}%
\bibitem [{\citenamefont {Yu}(2017)}]{yu_2017}%
  \BibitemOpen
  \bibfield  {author} {\bibinfo {author} {\bibfnamefont {S.}~\bibnamefont
  {Yu}},\ }\href@noop {} {\emph {\bibinfo {title} {Neuro-inspiring computing
  using resistive synaptic devices}}}\ (\bibinfo  {publisher} {Springer
  International Publishing},\ \bibinfo {year} {2017})\BibitemShut {NoStop}%
\bibitem [{\citenamefont {O'Reilly}\ \emph {et~al.}(2020)\citenamefont
  {O'Reilly}, \citenamefont {Munakata}, \citenamefont {Hazy},\ and\
  \citenamefont {Frank}}]{orrei2020}%
  \BibitemOpen
  \bibfield  {author} {\bibinfo {author} {\bibnamefont {O'Reilly}}, \bibinfo
  {author} {\bibnamefont {Munakata}}, \bibinfo {author} {\bibnamefont {Hazy}},
  \ and\ \bibinfo {author} {\bibnamefont {Frank}},\ }\href@noop {} {\emph
  {\bibinfo {title} {Computational Cognitive Neuroscience}}}\ (\bibinfo
  {publisher} {LibreTexts$^{TM}$},\ \bibinfo {year} {2020})\BibitemShut
  {NoStop}%
\bibitem [{\citenamefont {Prezioso}\ \emph {et~al.}(2016)\citenamefont
  {Prezioso}, \citenamefont {Merrikh~Bayat}, \citenamefont {Hoskins},
  \citenamefont {Likharev},\ and\ \citenamefont {Strukov}}]{prez16}%
  \BibitemOpen
  \bibfield  {author} {\bibinfo {author} {\bibfnamefont {M.}~\bibnamefont
  {Prezioso}}, \bibinfo {author} {\bibfnamefont {F.}~\bibnamefont
  {Merrikh~Bayat}}, \bibinfo {author} {\bibfnamefont {B.}~\bibnamefont
  {Hoskins}}, \bibinfo {author} {\bibfnamefont {K.}~\bibnamefont {Likharev}}, \
  and\ \bibinfo {author} {\bibfnamefont {D.}~\bibnamefont {Strukov}},\
  }\href@noop {} {\bibfield  {journal} {\bibinfo  {journal} {Sci. Reports}\
  }\textbf {\bibinfo {volume} {6}},\ \bibinfo {pages} {21331} (\bibinfo {year}
  {2016})}\BibitemShut {NoStop}%
\bibitem [{\citenamefont {Prezioso}\ \emph {et~al.}(2015)\citenamefont
  {Prezioso}, \citenamefont {Merrikh~Bayat}, \citenamefont {Hoskins},
  \citenamefont {Adam}, \citenamefont {Likharev},\ and\ \citenamefont
  {Strukov}}]{prez15}%
  \BibitemOpen
  \bibfield  {author} {\bibinfo {author} {\bibfnamefont {M.}~\bibnamefont
  {Prezioso}}, \bibinfo {author} {\bibfnamefont {F.}~\bibnamefont
  {Merrikh~Bayat}}, \bibinfo {author} {\bibfnamefont {B.~D.}\ \bibnamefont
  {Hoskins}}, \bibinfo {author} {\bibfnamefont {G.~C.}\ \bibnamefont {Adam}},
  \bibinfo {author} {\bibfnamefont {K.~K.}\ \bibnamefont {Likharev}}, \ and\
  \bibinfo {author} {\bibfnamefont {D.~B.}\ \bibnamefont {Strukov}},\
  }\href@noop {} {\bibfield  {journal} {\bibinfo  {journal} {Nature}\ }\textbf
  {\bibinfo {volume} {521}},\ \bibinfo {pages} {31} (\bibinfo {year}
  {2015})}\BibitemShut {NoStop}%
\bibitem [{\citenamefont {Sun}\ \emph {et~al.}(2019)\citenamefont {Sun},
  \citenamefont {Pedretti}, \citenamefont {Ambrosi}, \citenamefont {Bricalli},
  \citenamefont {Wang},\ and\ \citenamefont {Ielmini}}]{Sun4123}%
  \BibitemOpen
  \bibfield  {author} {\bibinfo {author} {\bibfnamefont {Z.}~\bibnamefont
  {Sun}}, \bibinfo {author} {\bibfnamefont {G.}~\bibnamefont {Pedretti}},
  \bibinfo {author} {\bibfnamefont {E.}~\bibnamefont {Ambrosi}}, \bibinfo
  {author} {\bibfnamefont {A.}~\bibnamefont {Bricalli}}, \bibinfo {author}
  {\bibfnamefont {W.}~\bibnamefont {Wang}}, \ and\ \bibinfo {author}
  {\bibfnamefont {D.}~\bibnamefont {Ielmini}},\ }\href@noop {} {\bibfield
  {journal} {\bibinfo  {journal} {Proceedings of the National Academy of
  Sciences}\ }\textbf {\bibinfo {volume} {116}},\ \bibinfo {pages} {4123}
  (\bibinfo {year} {2019})}\BibitemShut {NoStop}%
\bibitem [{\citenamefont {Sun}\ \emph {et~al.}(2020)\citenamefont {Sun},
  \citenamefont {Pedretti}, \citenamefont {Bricalli},\ and\ \citenamefont
  {Ielmini}}]{Sun2378}%
  \BibitemOpen
  \bibfield  {author} {\bibinfo {author} {\bibfnamefont {Z.}~\bibnamefont
  {Sun}}, \bibinfo {author} {\bibfnamefont {G.}~\bibnamefont {Pedretti}},
  \bibinfo {author} {\bibfnamefont {A.}~\bibnamefont {Bricalli}}, \ and\
  \bibinfo {author} {\bibfnamefont {D.}~\bibnamefont {Ielmini}},\ }\href@noop
  {} {\bibfield  {journal} {\bibinfo  {journal} {Science Adv.}\ }\textbf
  {\bibinfo {volume} {6}},\ \bibinfo {pages} {eaay2378} (\bibinfo {year}
  {2020})}\BibitemShut {NoStop}%
\bibitem [{\citenamefont {Gunkel}\ \emph {et~al.}(2020)\citenamefont {Gunkel},
  \citenamefont {Christensen}, \citenamefont {Chen},\ and\ \citenamefont
  {Pryds}}]{gun20}%
  \BibitemOpen
  \bibfield  {author} {\bibinfo {author} {\bibfnamefont {F.}~\bibnamefont
  {Gunkel}}, \bibinfo {author} {\bibfnamefont {D.~V.}\ \bibnamefont
  {Christensen}}, \bibinfo {author} {\bibfnamefont {Y.~Z.}\ \bibnamefont
  {Chen}}, \ and\ \bibinfo {author} {\bibfnamefont {N.}~\bibnamefont {Pryds}},\
  }\href@noop {} {\bibfield  {journal} {\bibinfo  {journal} {Appl. Phys.
  Lett.}\ }\textbf {\bibinfo {volume} {116}},\ \bibinfo {pages} {120505}
  (\bibinfo {year} {2020})}\BibitemShut {NoStop}%
\bibitem [{\citenamefont {Kwon}\ \emph {et~al.}(2010)\citenamefont {Kwon},
  \citenamefont {Kim}, \citenamefont {Jang}, \citenamefont {Jeon},
  \citenamefont {Lee}, \citenamefont {Kim}, \citenamefont {Li}, \citenamefont
  {Park}, \citenamefont {Lee}, \citenamefont {Han}, \citenamefont {Kim},\ and\
  \citenamefont {Hwang}}]{kwon10}%
  \BibitemOpen
  \bibfield  {author} {\bibinfo {author} {\bibfnamefont {D.-H.}\ \bibnamefont
  {Kwon}}, \bibinfo {author} {\bibfnamefont {K.~M.}\ \bibnamefont {Kim}},
  \bibinfo {author} {\bibfnamefont {J.~H.}\ \bibnamefont {Jang}}, \bibinfo
  {author} {\bibfnamefont {J.~M.}\ \bibnamefont {Jeon}}, \bibinfo {author}
  {\bibfnamefont {M.~H.}\ \bibnamefont {Lee}}, \bibinfo {author} {\bibfnamefont
  {G.~H.}\ \bibnamefont {Kim}}, \bibinfo {author} {\bibfnamefont {X.-S.}\
  \bibnamefont {Li}}, \bibinfo {author} {\bibfnamefont {G.-S.}\ \bibnamefont
  {Park}}, \bibinfo {author} {\bibfnamefont {B.}~\bibnamefont {Lee}}, \bibinfo
  {author} {\bibfnamefont {S.}~\bibnamefont {Han}}, \bibinfo {author}
  {\bibfnamefont {M.}~\bibnamefont {Kim}}, \ and\ \bibinfo {author}
  {\bibfnamefont {C.~S.}\ \bibnamefont {Hwang}},\ }\href@noop {} {\bibfield
  {journal} {\bibinfo  {journal} {Nature Nanotech.}\ }\textbf {\bibinfo
  {volume} {5}},\ \bibinfo {pages} {148} (\bibinfo {year} {2010})}\BibitemShut
  {NoStop}%
\bibitem [{\citenamefont {Herpers}\ \emph {et~al.}(2014)\citenamefont
  {Herpers}, \citenamefont {Lenser}, \citenamefont {Park}, \citenamefont
  {Offi}, \citenamefont {Borgatti}, \citenamefont {Panaccione}, \citenamefont
  {Menzel}, \citenamefont {Waser},\ and\ \citenamefont {Dittmann}}]{herpers}%
  \BibitemOpen
  \bibfield  {author} {\bibinfo {author} {\bibfnamefont {A.}~\bibnamefont
  {Herpers}}, \bibinfo {author} {\bibfnamefont {C.}~\bibnamefont {Lenser}},
  \bibinfo {author} {\bibfnamefont {C.}~\bibnamefont {Park}}, \bibinfo {author}
  {\bibfnamefont {F.}~\bibnamefont {Offi}}, \bibinfo {author} {\bibfnamefont
  {F.}~\bibnamefont {Borgatti}}, \bibinfo {author} {\bibfnamefont
  {G.}~\bibnamefont {Panaccione}}, \bibinfo {author} {\bibfnamefont
  {S.}~\bibnamefont {Menzel}}, \bibinfo {author} {\bibfnamefont
  {R.}~\bibnamefont {Waser}}, \ and\ \bibinfo {author} {\bibfnamefont
  {R.}~\bibnamefont {Dittmann}},\ }\href@noop {} {\bibfield  {journal}
  {\bibinfo  {journal} {Advanced Materials}\ }\textbf {\bibinfo {volume}
  {26}},\ \bibinfo {pages} {2730} (\bibinfo {year} {2014})}\BibitemShut
  {NoStop}%
\bibitem [{\citenamefont {Nian}\ \emph {et~al.}(2007)\citenamefont {Nian},
  \citenamefont {Strozier}, \citenamefont {Wu}, \citenamefont {Chen},\ and\
  \citenamefont {Ignatiev}}]{nian2007}%
  \BibitemOpen
  \bibfield  {author} {\bibinfo {author} {\bibfnamefont {Y.~B.}\ \bibnamefont
  {Nian}}, \bibinfo {author} {\bibfnamefont {J.}~\bibnamefont {Strozier}},
  \bibinfo {author} {\bibfnamefont {N.~J.}\ \bibnamefont {Wu}}, \bibinfo
  {author} {\bibfnamefont {X.}~\bibnamefont {Chen}}, \ and\ \bibinfo {author}
  {\bibfnamefont {A.}~\bibnamefont {Ignatiev}},\ }\href@noop {} {\bibfield
  {journal} {\bibinfo  {journal} {Phys. Rev. Lett.}\ }\textbf {\bibinfo
  {volume} {98}},\ \bibinfo {pages} {146403} (\bibinfo {year}
  {2007})}\BibitemShut {NoStop}%
\bibitem [{\citenamefont {Rozenberg}\ \emph {et~al.}(2010)\citenamefont
  {Rozenberg}, \citenamefont {S\'anchez}, \citenamefont {Weht}, \citenamefont
  {Acha}, \citenamefont {Gomez-Marlasca},\ and\ \citenamefont
  {Levy}}]{rozenberg_2010}%
  \BibitemOpen
  \bibfield  {author} {\bibinfo {author} {\bibfnamefont {M.~J.}\ \bibnamefont
  {Rozenberg}}, \bibinfo {author} {\bibfnamefont {M.~J.}\ \bibnamefont
  {S\'anchez}}, \bibinfo {author} {\bibfnamefont {R.}~\bibnamefont {Weht}},
  \bibinfo {author} {\bibfnamefont {C.}~\bibnamefont {Acha}}, \bibinfo {author}
  {\bibfnamefont {F.}~\bibnamefont {Gomez-Marlasca}}, \ and\ \bibinfo {author}
  {\bibfnamefont {P.}~\bibnamefont {Levy}},\ }\href@noop {} {\bibfield
  {journal} {\bibinfo  {journal} {Phys. Rev. B}\ }\textbf {\bibinfo {volume}
  {81}},\ \bibinfo {pages} {115101} (\bibinfo {year} {2010})}\BibitemShut
  {NoStop}%
\bibitem [{\citenamefont {Acharya}\ \emph {et~al.}(2017)\citenamefont
  {Acharya}, \citenamefont {Jo}, \citenamefont {Raveendra}, \citenamefont
  {Dash}, \citenamefont {Kim}, \citenamefont {Baik}, \citenamefont {Lee},
  \citenamefont {Park}, \citenamefont {Lee}, \citenamefont {Chae},
  \citenamefont {Hwang},\ and\ \citenamefont {Jung}}]{acharya2017}%
  \BibitemOpen
  \bibfield  {author} {\bibinfo {author} {\bibfnamefont {S.~K.}\ \bibnamefont
  {Acharya}}, \bibinfo {author} {\bibfnamefont {J.}~\bibnamefont {Jo}},
  \bibinfo {author} {\bibfnamefont {N.~V.}\ \bibnamefont {Raveendra}}, \bibinfo
  {author} {\bibfnamefont {U.}~\bibnamefont {Dash}}, \bibinfo {author}
  {\bibfnamefont {M.}~\bibnamefont {Kim}}, \bibinfo {author} {\bibfnamefont
  {H.}~\bibnamefont {Baik}}, \bibinfo {author} {\bibfnamefont {S.}~\bibnamefont
  {Lee}}, \bibinfo {author} {\bibfnamefont {B.~H.}\ \bibnamefont {Park}},
  \bibinfo {author} {\bibfnamefont {J.~S.}\ \bibnamefont {Lee}}, \bibinfo
  {author} {\bibfnamefont {S.~C.}\ \bibnamefont {Chae}}, \bibinfo {author}
  {\bibfnamefont {C.~S.}\ \bibnamefont {Hwang}}, \ and\ \bibinfo {author}
  {\bibfnamefont {C.~U.}\ \bibnamefont {Jung}},\ }\href@noop {} {\bibfield
  {journal} {\bibinfo  {journal} {Nanoscale}\ }\textbf {\bibinfo {volume}
  {9}},\ \bibinfo {pages} {10502} (\bibinfo {year} {2017})}\BibitemShut
  {NoStop}%
\bibitem [{\citenamefont {Nallagatla}\ \emph
  {et~al.}(2019{\natexlab{a}})\citenamefont {Nallagatla}, \citenamefont {Jo},
  \citenamefont {Acharya}, \citenamefont {Kim},\ and\ \citenamefont
  {Jung}}]{nall2019}%
  \BibitemOpen
  \bibfield  {author} {\bibinfo {author} {\bibfnamefont {V.~R.}\ \bibnamefont
  {Nallagatla}}, \bibinfo {author} {\bibfnamefont {J.}~\bibnamefont {Jo}},
  \bibinfo {author} {\bibfnamefont {S.~K.}\ \bibnamefont {Acharya}}, \bibinfo
  {author} {\bibfnamefont {M.}~\bibnamefont {Kim}}, \ and\ \bibinfo {author}
  {\bibfnamefont {C.~U.}\ \bibnamefont {Jung}},\ }\href@noop {} {\bibfield
  {journal} {\bibinfo  {journal} {Sci. Reports}\ }\textbf {\bibinfo {volume}
  {9}},\ \bibinfo {pages} {1188} (\bibinfo {year}
  {2019}{\natexlab{a}})}\BibitemShut {NoStop}%
\bibitem [{\citenamefont {Lo}\ \emph {et~al.}(2020)\citenamefont {Lo},
  \citenamefont {Yang}, \citenamefont {Huang}, \citenamefont {Huang},
  \citenamefont {Chen}, \citenamefont {Huang}, \citenamefont {Chu},\ and\
  \citenamefont {Wu}}]{hung2020}%
  \BibitemOpen
  \bibfield  {author} {\bibinfo {author} {\bibfnamefont {H.-Y.}\ \bibnamefont
  {Lo}}, \bibinfo {author} {\bibfnamefont {C.-Y.}\ \bibnamefont {Yang}},
  \bibinfo {author} {\bibfnamefont {G.-M.}\ \bibnamefont {Huang}}, \bibinfo
  {author} {\bibfnamefont {C.-Y.}\ \bibnamefont {Huang}}, \bibinfo {author}
  {\bibfnamefont {J.-Y.}\ \bibnamefont {Chen}}, \bibinfo {author}
  {\bibfnamefont {C.-W.}\ \bibnamefont {Huang}}, \bibinfo {author}
  {\bibfnamefont {Y.-H.}\ \bibnamefont {Chu}}, \ and\ \bibinfo {author}
  {\bibfnamefont {W.-W.}\ \bibnamefont {Wu}},\ }\href@noop {} {\bibfield
  {journal} {\bibinfo  {journal} {Nano Energy}\ }\textbf {\bibinfo {volume}
  {72}},\ \bibinfo {pages} {104683} (\bibinfo {year} {2020})}\BibitemShut
  {NoStop}%
\bibitem [{\citenamefont {Mou}\ \emph {et~al.}(2021)\citenamefont {Mou},
  \citenamefont {Tang}, \citenamefont {Lyu}, \citenamefont {Zhang},
  \citenamefont {Yang}, \citenamefont {Xu}, \citenamefont {Liu}, \citenamefont
  {Xu}, \citenamefont {Zhou}, \citenamefont {Sun}, \citenamefont {Zhong},
  \citenamefont {Gao}, \citenamefont {Yu}, \citenamefont {Qian},\ and\
  \citenamefont {Wu}}]{mou2021}%
  \BibitemOpen
  \bibfield  {author} {\bibinfo {author} {\bibfnamefont {X.}~\bibnamefont
  {Mou}}, \bibinfo {author} {\bibfnamefont {J.}~\bibnamefont {Tang}}, \bibinfo
  {author} {\bibfnamefont {Y.}~\bibnamefont {Lyu}}, \bibinfo {author}
  {\bibfnamefont {Q.}~\bibnamefont {Zhang}}, \bibinfo {author} {\bibfnamefont
  {S.}~\bibnamefont {Yang}}, \bibinfo {author} {\bibfnamefont {F.}~\bibnamefont
  {Xu}}, \bibinfo {author} {\bibfnamefont {W.}~\bibnamefont {Liu}}, \bibinfo
  {author} {\bibfnamefont {M.}~\bibnamefont {Xu}}, \bibinfo {author}
  {\bibfnamefont {Y.}~\bibnamefont {Zhou}}, \bibinfo {author} {\bibfnamefont
  {W.}~\bibnamefont {Sun}}, \bibinfo {author} {\bibfnamefont {Y.}~\bibnamefont
  {Zhong}}, \bibinfo {author} {\bibfnamefont {B.}~\bibnamefont {Gao}}, \bibinfo
  {author} {\bibfnamefont {P.}~\bibnamefont {Yu}}, \bibinfo {author}
  {\bibfnamefont {H.}~\bibnamefont {Qian}}, \ and\ \bibinfo {author}
  {\bibfnamefont {H.}~\bibnamefont {Wu}},\ }\href@noop {} {\bibfield  {journal}
  {\bibinfo  {journal} {Science Adv.}\ }\textbf {\bibinfo {volume} {7}},\
  \bibinfo {pages} {eabh0648} (\bibinfo {year} {2021})}\BibitemShut {NoStop}%
\bibitem [{\citenamefont {Nallagatla}\ \emph
  {et~al.}(2019{\natexlab{b}})\citenamefont {Nallagatla}, \citenamefont
  {Heisig}, \citenamefont {Baeumer}, \citenamefont {Feyer}, \citenamefont
  {Jugovac}, \citenamefont {Zamborlini}, \citenamefont {Schneider},
  \citenamefont {Waser}, \citenamefont {Kim}, \citenamefont {Jung},\ and\
  \citenamefont {Dittmann}}]{nalla2019}%
  \BibitemOpen
  \bibfield  {author} {\bibinfo {author} {\bibfnamefont {V.~R.}\ \bibnamefont
  {Nallagatla}}, \bibinfo {author} {\bibfnamefont {T.}~\bibnamefont {Heisig}},
  \bibinfo {author} {\bibfnamefont {C.}~\bibnamefont {Baeumer}}, \bibinfo
  {author} {\bibfnamefont {V.}~\bibnamefont {Feyer}}, \bibinfo {author}
  {\bibfnamefont {M.}~\bibnamefont {Jugovac}}, \bibinfo {author} {\bibfnamefont
  {G.}~\bibnamefont {Zamborlini}}, \bibinfo {author} {\bibfnamefont {C.~M.}\
  \bibnamefont {Schneider}}, \bibinfo {author} {\bibfnamefont {R.}~\bibnamefont
  {Waser}}, \bibinfo {author} {\bibfnamefont {M.}~\bibnamefont {Kim}}, \bibinfo
  {author} {\bibfnamefont {C.~U.}\ \bibnamefont {Jung}}, \ and\ \bibinfo
  {author} {\bibfnamefont {R.}~\bibnamefont {Dittmann}},\ }\href@noop {}
  {\bibfield  {journal} {\bibinfo  {journal} {Advanced Materials}\ }\textbf
  {\bibinfo {volume} {31}},\ \bibinfo {pages} {1903391} (\bibinfo {year}
  {2019}{\natexlab{b}})}\BibitemShut {NoStop}%
\bibitem [{\citenamefont {Nallagatla}\ and\ \citenamefont
  {Jung}(2020)}]{nalla2020}%
  \BibitemOpen
  \bibfield  {author} {\bibinfo {author} {\bibfnamefont {V.~R.}\ \bibnamefont
  {Nallagatla}}\ and\ \bibinfo {author} {\bibfnamefont {C.~U.}\ \bibnamefont
  {Jung}},\ }\href@noop {} {\bibfield  {journal} {\bibinfo  {journal} {Applied
  Physics Letters}\ }\textbf {\bibinfo {volume} {117}},\ \bibinfo {pages}
  {143503} (\bibinfo {year} {2020})}\BibitemShut {NoStop}%
\bibitem [{\citenamefont {Kim}\ \emph {et~al.}(2020)\citenamefont {Kim},
  \citenamefont {Nallagatla}, \citenamefont {Kwon}, \citenamefont {Jung},\ and\
  \citenamefont {Kim}}]{kim2020}%
  \BibitemOpen
  \bibfield  {author} {\bibinfo {author} {\bibfnamefont {H.~G.}\ \bibnamefont
  {Kim}}, \bibinfo {author} {\bibfnamefont {V.~R.}\ \bibnamefont {Nallagatla}},
  \bibinfo {author} {\bibfnamefont {D.-H.}\ \bibnamefont {Kwon}}, \bibinfo
  {author} {\bibfnamefont {C.~U.}\ \bibnamefont {Jung}}, \ and\ \bibinfo
  {author} {\bibfnamefont {M.}~\bibnamefont {Kim}},\ }\href@noop {} {\bibfield
  {journal} {\bibinfo  {journal} {Journal of Applied Physics}\ }\textbf
  {\bibinfo {volume} {128}},\ \bibinfo {pages} {074501} (\bibinfo {year}
  {2020})}\BibitemShut {NoStop}%
\bibitem [{\citenamefont {Yao}\ \emph {et~al.}(2020)\citenamefont {Yao},
  \citenamefont {Inkinen},\ and\ \citenamefont {van Dijken}}]{yao20}%
  \BibitemOpen
  \bibfield  {author} {\bibinfo {author} {\bibfnamefont {L.}~\bibnamefont
  {Yao}}, \bibinfo {author} {\bibfnamefont {S.}~\bibnamefont {Inkinen}}, \ and\
  \bibinfo {author} {\bibfnamefont {S.}~\bibnamefont {van Dijken}},\
  }\href@noop {} {\bibfield  {journal} {\bibinfo  {journal} {Appl. Phys.
  Lett.}\ }\textbf {\bibinfo {volume} {116}},\ \bibinfo {pages} {120505}
  (\bibinfo {year} {2020})}\BibitemShut {NoStop}%
\bibitem [{\citenamefont {Cho}\ \emph {et~al.}(2016)\citenamefont {Cho},
  \citenamefont {Yun}, \citenamefont {Tappertzhofen}, \citenamefont
  {Kursumovic}, \citenamefont {Lee}, \citenamefont {Lu}, \citenamefont {Jia},
  \citenamefont {Fan}, \citenamefont {Jian}, \citenamefont {Wang},
  \citenamefont {Hofmann},\ and\ \citenamefont {MacManus-Driscoll}}]{cho2016}%
  \BibitemOpen
  \bibfield  {author} {\bibinfo {author} {\bibfnamefont {S.}~\bibnamefont
  {Cho}}, \bibinfo {author} {\bibfnamefont {C.}~\bibnamefont {Yun}}, \bibinfo
  {author} {\bibfnamefont {S.}~\bibnamefont {Tappertzhofen}}, \bibinfo {author}
  {\bibfnamefont {A.}~\bibnamefont {Kursumovic}}, \bibinfo {author}
  {\bibfnamefont {S.}~\bibnamefont {Lee}}, \bibinfo {author} {\bibfnamefont
  {P.}~\bibnamefont {Lu}}, \bibinfo {author} {\bibfnamefont {Q.}~\bibnamefont
  {Jia}}, \bibinfo {author} {\bibfnamefont {M.}~\bibnamefont {Fan}}, \bibinfo
  {author} {\bibfnamefont {J.}~\bibnamefont {Jian}}, \bibinfo {author}
  {\bibfnamefont {H.}~\bibnamefont {Wang}}, \bibinfo {author} {\bibfnamefont
  {S.}~\bibnamefont {Hofmann}}, \ and\ \bibinfo {author} {\bibfnamefont
  {J.~L.}\ \bibnamefont {MacManus-Driscoll}},\ }\href@noop {} {\bibfield
  {journal} {\bibinfo  {journal} {Nature Commun.}\ }\textbf {\bibinfo {volume}
  {7}},\ \bibinfo {pages} {12373} (\bibinfo {year} {2016})}\BibitemShut
  {NoStop}%
\bibitem [{\citenamefont {Aguadero}\ \emph {et~al.}(2011)\citenamefont
  {Aguadero}, \citenamefont {Falcon}, \citenamefont {Campos-Martin},
  \citenamefont {Al-Zahrani}, \citenamefont {Fierro},\ and\ \citenamefont
  {Alonso}}]{agua2011}%
  \BibitemOpen
  \bibfield  {author} {\bibinfo {author} {\bibfnamefont {A.}~\bibnamefont
  {Aguadero}}, \bibinfo {author} {\bibfnamefont {H.}~\bibnamefont {Falcon}},
  \bibinfo {author} {\bibfnamefont {J.~M.}\ \bibnamefont {Campos-Martin}},
  \bibinfo {author} {\bibfnamefont {S.~M.}\ \bibnamefont {Al-Zahrani}},
  \bibinfo {author} {\bibfnamefont {J.~L.~G.}\ \bibnamefont {Fierro}}, \ and\
  \bibinfo {author} {\bibfnamefont {J.~A.}\ \bibnamefont {Alonso}},\
  }\href@noop {} {\bibfield  {journal} {\bibinfo  {journal} {Angewandte Chemie
  International Edition}\ }\textbf {\bibinfo {volume} {50}},\ \bibinfo {pages}
  {6557} (\bibinfo {year} {2011})}\BibitemShut {NoStop}%
\bibitem [{\citenamefont {Rom\'an~Acevedo}\ \emph {et~al.}(2020)\citenamefont
  {Rom\'an~Acevedo}, \citenamefont {van~den Bosch}, \citenamefont {Aguirre},
  \citenamefont {Acha}, \citenamefont {Cavallaro}, \citenamefont {Ferreyra},
  \citenamefont {S\'anchez}, \citenamefont {Patrone}, \citenamefont
  {Aguadero},\ and\ \citenamefont {Rubi}}]{roman2020}%
  \BibitemOpen
  \bibfield  {author} {\bibinfo {author} {\bibfnamefont {W.}~\bibnamefont
  {Rom\'an~Acevedo}}, \bibinfo {author} {\bibfnamefont {C.~A.~M.}\ \bibnamefont
  {van~den Bosch}}, \bibinfo {author} {\bibfnamefont {M.~H.}\ \bibnamefont
  {Aguirre}}, \bibinfo {author} {\bibfnamefont {C.}~\bibnamefont {Acha}},
  \bibinfo {author} {\bibfnamefont {A.}~\bibnamefont {Cavallaro}}, \bibinfo
  {author} {\bibfnamefont {C.}~\bibnamefont {Ferreyra}}, \bibinfo {author}
  {\bibfnamefont {M.~J.}\ \bibnamefont {S\'anchez}}, \bibinfo {author}
  {\bibfnamefont {L.}~\bibnamefont {Patrone}}, \bibinfo {author} {\bibfnamefont
  {A.}~\bibnamefont {Aguadero}}, \ and\ \bibinfo {author} {\bibfnamefont
  {D.}~\bibnamefont {Rubi}},\ }\href@noop {} {\bibfield  {journal} {\bibinfo
  {journal} {Applied Physics Letters}\ }\textbf {\bibinfo {volume} {116}},\
  \bibinfo {pages} {063502} (\bibinfo {year} {2020})}\BibitemShut {NoStop}%
\bibitem [{\citenamefont {Wu}\ \emph {et~al.}(2011)\citenamefont {Wu},
  \citenamefont {Peng},\ and\ \citenamefont {Wu}}]{wu2011}%
  \BibitemOpen
  \bibfield  {author} {\bibinfo {author} {\bibfnamefont {S.~X.}\ \bibnamefont
  {Wu}}, \bibinfo {author} {\bibfnamefont {H.~Y.}\ \bibnamefont {Peng}}, \ and\
  \bibinfo {author} {\bibfnamefont {T.}~\bibnamefont {Wu}},\ }\href@noop {}
  {\bibfield  {journal} {\bibinfo  {journal} {Applied Physics Letters}\
  }\textbf {\bibinfo {volume} {98}},\ \bibinfo {pages} {093503} (\bibinfo
  {year} {2011})}\BibitemShut {NoStop}%
\bibitem [{\citenamefont {Salaoru}\ \emph {et~al.}(2013)\citenamefont
  {Salaoru}, \citenamefont {Khiat}, \citenamefont {Li}, \citenamefont
  {Berdan},\ and\ \citenamefont {Prodromakis}}]{sala2013}%
  \BibitemOpen
  \bibfield  {author} {\bibinfo {author} {\bibfnamefont {I.}~\bibnamefont
  {Salaoru}}, \bibinfo {author} {\bibfnamefont {A.}~\bibnamefont {Khiat}},
  \bibinfo {author} {\bibfnamefont {Q.}~\bibnamefont {Li}}, \bibinfo {author}
  {\bibfnamefont {R.}~\bibnamefont {Berdan}}, \ and\ \bibinfo {author}
  {\bibfnamefont {T.}~\bibnamefont {Prodromakis}},\ }\href@noop {} {\bibfield
  {journal} {\bibinfo  {journal} {Applied Physics Letters}\ }\textbf {\bibinfo
  {volume} {103}},\ \bibinfo {pages} {233513} (\bibinfo {year}
  {2013})}\BibitemShut {NoStop}%
\bibitem [{\citenamefont {Salaoru}\ \emph {et~al.}(2014)\citenamefont
  {Salaoru}, \citenamefont {Li}, \citenamefont {Khiat},\ and\ \citenamefont
  {Prodromakis}}]{salarou2014}%
  \BibitemOpen
  \bibfield  {author} {\bibinfo {author} {\bibfnamefont {I.}~\bibnamefont
  {Salaoru}}, \bibinfo {author} {\bibfnamefont {Q.}~\bibnamefont {Li}},
  \bibinfo {author} {\bibfnamefont {A.}~\bibnamefont {Khiat}}, \ and\ \bibinfo
  {author} {\bibfnamefont {T.}~\bibnamefont {Prodromakis}},\ }\href@noop {}
  {\bibfield  {journal} {\bibinfo  {journal} {Nanoscale Research Letters}\
  }\textbf {\bibinfo {volume} {9}},\ \bibinfo {pages} {552} (\bibinfo {year}
  {2014})}\BibitemShut {NoStop}%
\bibitem [{\citenamefont {Bessonov}\ \emph {et~al.}(2015)\citenamefont
  {Bessonov}, \citenamefont {Kirikova}, \citenamefont {Petukhov}, \citenamefont
  {Allen}, \citenamefont {Ryhänen},\ and\ \citenamefont {Bailey}}]{besso15}%
  \BibitemOpen
  \bibfield  {author} {\bibinfo {author} {\bibfnamefont {A.~A.}\ \bibnamefont
  {Bessonov}}, \bibinfo {author} {\bibfnamefont {M.~N.}\ \bibnamefont
  {Kirikova}}, \bibinfo {author} {\bibfnamefont {D.~I.}\ \bibnamefont
  {Petukhov}}, \bibinfo {author} {\bibfnamefont {M.}~\bibnamefont {Allen}},
  \bibinfo {author} {\bibfnamefont {T.}~\bibnamefont {Ryhänen}}, \ and\
  \bibinfo {author} {\bibfnamefont {M.~J.~A.}\ \bibnamefont {Bailey}},\
  }\href@noop {} {\bibfield  {journal} {\bibinfo  {journal} {Nature Mater.}\
  }\textbf {\bibinfo {volume} {14}},\ \bibinfo {pages} {199} (\bibinfo {year}
  {2015})}\BibitemShut {NoStop}%
\bibitem [{\citenamefont {Yang}\ \emph {et~al.}(2017)\citenamefont {Yang},
  \citenamefont {Kim}, \citenamefont {Zheng}, \citenamefont {Beom},
  \citenamefont {Park}, \citenamefont {Kang},\ and\ \citenamefont
  {Yoon}}]{Yang_2017}%
  \BibitemOpen
  \bibfield  {author} {\bibinfo {author} {\bibfnamefont {P.}~\bibnamefont
  {Yang}}, \bibinfo {author} {\bibfnamefont {H.~J.}\ \bibnamefont {Kim}},
  \bibinfo {author} {\bibfnamefont {H.}~\bibnamefont {Zheng}}, \bibinfo
  {author} {\bibfnamefont {G.~W.}\ \bibnamefont {Beom}}, \bibinfo {author}
  {\bibfnamefont {J.-S.}\ \bibnamefont {Park}}, \bibinfo {author}
  {\bibfnamefont {C.~J.}\ \bibnamefont {Kang}}, \ and\ \bibinfo {author}
  {\bibfnamefont {T.-S.}\ \bibnamefont {Yoon}},\ }\href@noop {} {\bibfield
  {journal} {\bibinfo  {journal} {Nanotechnology}\ }\textbf {\bibinfo {volume}
  {28}},\ \bibinfo {pages} {225201} (\bibinfo {year} {2017})}\BibitemShut
  {NoStop}%
\bibitem [{\citenamefont {Park}\ \emph {et~al.}(2018)\citenamefont {Park},
  \citenamefont {Yang}, \citenamefont {Kim}, \citenamefont {Beom},
  \citenamefont {Lee}, \citenamefont {Kang},\ and\ \citenamefont
  {Yoon}}]{park2018}%
  \BibitemOpen
  \bibfield  {author} {\bibinfo {author} {\bibfnamefont {D.}~\bibnamefont
  {Park}}, \bibinfo {author} {\bibfnamefont {P.}~\bibnamefont {Yang}}, \bibinfo
  {author} {\bibfnamefont {H.~J.}\ \bibnamefont {Kim}}, \bibinfo {author}
  {\bibfnamefont {K.}~\bibnamefont {Beom}}, \bibinfo {author} {\bibfnamefont
  {H.~H.}\ \bibnamefont {Lee}}, \bibinfo {author} {\bibfnamefont {C.~J.}\
  \bibnamefont {Kang}}, \ and\ \bibinfo {author} {\bibfnamefont {T.-S.}\
  \bibnamefont {Yoon}},\ }\href@noop {} {\bibfield  {journal} {\bibinfo
  {journal} {Applied Physics Letters}\ }\textbf {\bibinfo {volume} {113}},\
  \bibinfo {pages} {162102} (\bibinfo {year} {2018})}\BibitemShut {NoStop}%
\bibitem [{\citenamefont {Liu}\ \emph {et~al.}(2018)\citenamefont {Liu},
  \citenamefont {Dong}, \citenamefont {Yan}, \citenamefont {Yuan},
  \citenamefont {Zhang}, \citenamefont {Yang},\ and\ \citenamefont
  {Xiao}}]{Liu_2018}%
  \BibitemOpen
  \bibfield  {author} {\bibinfo {author} {\bibfnamefont {R.}~\bibnamefont
  {Liu}}, \bibinfo {author} {\bibfnamefont {R.}~\bibnamefont {Dong}}, \bibinfo
  {author} {\bibfnamefont {X.}~\bibnamefont {Yan}}, \bibinfo {author}
  {\bibfnamefont {S.}~\bibnamefont {Yuan}}, \bibinfo {author} {\bibfnamefont
  {D.}~\bibnamefont {Zhang}}, \bibinfo {author} {\bibfnamefont
  {B.}~\bibnamefont {Yang}}, \ and\ \bibinfo {author} {\bibfnamefont
  {X.}~\bibnamefont {Xiao}},\ }\href@noop {} {\bibfield  {journal} {\bibinfo
  {journal} {Applied Physics Express}\ }\textbf {\bibinfo {volume} {11}},\
  \bibinfo {pages} {114103} (\bibinfo {year} {2018})}\BibitemShut {NoStop}%
\bibitem [{\citenamefont {Borowiec}\ \emph {et~al.}(2020)\citenamefont
  {Borowiec}, \citenamefont {Liu}, \citenamefont {Liang}, \citenamefont
  {Kreouzis}, \citenamefont {Bevan}, \citenamefont {He}, \citenamefont {Ma},\
  and\ \citenamefont {Gillin}}]{boro2020}%
  \BibitemOpen
  \bibfield  {author} {\bibinfo {author} {\bibfnamefont {J.}~\bibnamefont
  {Borowiec}}, \bibinfo {author} {\bibfnamefont {M.}~\bibnamefont {Liu}},
  \bibinfo {author} {\bibfnamefont {W.}~\bibnamefont {Liang}}, \bibinfo
  {author} {\bibfnamefont {T.}~\bibnamefont {Kreouzis}}, \bibinfo {author}
  {\bibfnamefont {A.~J.}\ \bibnamefont {Bevan}}, \bibinfo {author}
  {\bibfnamefont {Y.}~\bibnamefont {He}}, \bibinfo {author} {\bibfnamefont
  {Y.}~\bibnamefont {Ma}}, \ and\ \bibinfo {author} {\bibfnamefont {W.~P.}\
  \bibnamefont {Gillin}},\ }\href@noop {} {\bibfield  {journal} {\bibinfo
  {journal} {Nanomaterials}\ }\textbf {\bibinfo {volume} {10}},\ \bibinfo
  {pages} {2103} (\bibinfo {year} {2020})}\BibitemShut {NoStop}%
\bibitem [{\citenamefont {Guo}\ \emph {et~al.}(2020)\citenamefont {Guo},
  \citenamefont {Huang}, \citenamefont {Zhou}, \citenamefont {Chang},
  \citenamefont {Cao}, \citenamefont {Xiao},\ and\ \citenamefont
  {Shi}}]{guo2020}%
  \BibitemOpen
  \bibfield  {author} {\bibinfo {author} {\bibfnamefont {X.}~\bibnamefont
  {Guo}}, \bibinfo {author} {\bibfnamefont {L.}~\bibnamefont {Huang}}, \bibinfo
  {author} {\bibfnamefont {X.}~\bibnamefont {Zhou}}, \bibinfo {author}
  {\bibfnamefont {Q.}~\bibnamefont {Chang}}, \bibinfo {author} {\bibfnamefont
  {C.}~\bibnamefont {Cao}}, \bibinfo {author} {\bibfnamefont {G.}~\bibnamefont
  {Xiao}}, \ and\ \bibinfo {author} {\bibfnamefont {W.}~\bibnamefont {Shi}},\
  }\href@noop {} {\bibfield  {journal} {\bibinfo  {journal} {Advanced
  Functional Materials}\ }\textbf {\bibinfo {volume} {30}},\ \bibinfo {pages}
  {2003635} (\bibinfo {year} {2020})}\BibitemShut {NoStop}%
\bibitem [{\citenamefont {Tran}\ and\ \citenamefont
  {Teuscher}(2017)}]{tran2017}%
  \BibitemOpen
  \bibfield  {author} {\bibinfo {author} {\bibfnamefont {S.~J.~D.}\
  \bibnamefont {Tran}}\ and\ \bibinfo {author} {\bibfnamefont {C.}~\bibnamefont
  {Teuscher}},\ }\href@noop {} {\bibfield  {journal} {\bibinfo  {journal} {Int.
  J. Unconv. Comput.}\ }\textbf {\bibinfo {volume} {13}},\ \bibinfo {pages}
  {35} (\bibinfo {year} {2017})}\BibitemShut {NoStop}%
\bibitem [{\citenamefont {Tomio}\ \emph {et~al.}(1994)\citenamefont {Tomio},
  \citenamefont {Miki}, \citenamefont {Tabata}, \citenamefont {Kawai},\ and\
  \citenamefont {Kawai}}]{tomio94}%
  \BibitemOpen
  \bibfield  {author} {\bibinfo {author} {\bibfnamefont {T.}~\bibnamefont
  {Tomio}}, \bibinfo {author} {\bibfnamefont {H.}~\bibnamefont {Miki}},
  \bibinfo {author} {\bibfnamefont {H.}~\bibnamefont {Tabata}}, \bibinfo
  {author} {\bibfnamefont {T.}~\bibnamefont {Kawai}}, \ and\ \bibinfo {author}
  {\bibfnamefont {S.}~\bibnamefont {Kawai}},\ }\href@noop {} {\bibfield
  {journal} {\bibinfo  {journal} {Journal of Applied Physics}\ }\textbf
  {\bibinfo {volume} {76}},\ \bibinfo {pages} {5886} (\bibinfo {year}
  {1994})}\BibitemShut {NoStop}%
\bibitem [{\citenamefont {Catalan}(2006)}]{catalan2006}%
  \BibitemOpen
  \bibfield  {author} {\bibinfo {author} {\bibfnamefont {G.}~\bibnamefont
  {Catalan}},\ }\href@noop {} {\bibfield  {journal} {\bibinfo  {journal}
  {Applied Physics Letters}\ }\textbf {\bibinfo {volume} {88}},\ \bibinfo
  {pages} {102902} (\bibinfo {year} {2006})}\BibitemShut {NoStop}%
\bibitem [{\citenamefont {Sayed}\ \emph {et~al.}(2014)\citenamefont {Sayed},
  \citenamefont {Achary}, \citenamefont {Deshpande}, \citenamefont {Rajeswari},
  \citenamefont {Kadam}, \citenamefont {Dwebedi}, \citenamefont {Nigam},\ and\
  \citenamefont {Tyagi}}]{sayed2014}%
  \BibitemOpen
  \bibfield  {author} {\bibinfo {author} {\bibfnamefont {F.~N.}\ \bibnamefont
  {Sayed}}, \bibinfo {author} {\bibfnamefont {S.~N.}\ \bibnamefont {Achary}},
  \bibinfo {author} {\bibfnamefont {S.}~\bibnamefont {Deshpande}}, \bibinfo
  {author} {\bibfnamefont {B.}~\bibnamefont {Rajeswari}}, \bibinfo {author}
  {\bibfnamefont {R.~M.}\ \bibnamefont {Kadam}}, \bibinfo {author}
  {\bibfnamefont {S.}~\bibnamefont {Dwebedi}}, \bibinfo {author} {\bibfnamefont
  {A.~K.}\ \bibnamefont {Nigam}}, \ and\ \bibinfo {author} {\bibfnamefont
  {A.~K.}\ \bibnamefont {Tyagi}},\ }\href@noop {} {\bibfield  {journal}
  {\bibinfo  {journal} {Zeitschrift für anorganische und allgemeine Chemie}\
  }\textbf {\bibinfo {volume} {640}},\ \bibinfo {pages} {1907} (\bibinfo {year}
  {2014})}\BibitemShut {NoStop}%
\bibitem [{\citenamefont {Ghenzi}\ \emph {et~al.}(2013)\citenamefont {Ghenzi},
  \citenamefont {S{\'{a}}nchez},\ and\ \citenamefont {Levy}}]{ghenzi_2013}%
  \BibitemOpen
  \bibfield  {author} {\bibinfo {author} {\bibfnamefont {N.}~\bibnamefont
  {Ghenzi}}, \bibinfo {author} {\bibfnamefont {M.~J.}\ \bibnamefont
  {S{\'{a}}nchez}}, \ and\ \bibinfo {author} {\bibfnamefont {P.}~\bibnamefont
  {Levy}},\ }\href@noop {} {\bibfield  {journal} {\bibinfo  {journal} {Journal
  of Physics D: Applied Physics}\ }\textbf {\bibinfo {volume} {46}},\ \bibinfo
  {pages} {415101} (\bibinfo {year} {2013})}\BibitemShut {NoStop}%
\bibitem [{\citenamefont {Spinelli}\ \emph {et~al.}(2010)\citenamefont
  {Spinelli}, \citenamefont {Torija}, \citenamefont {Liu}, \citenamefont
  {Jan},\ and\ \citenamefont {Leighton}}]{spinelli2010}%
  \BibitemOpen
  \bibfield  {author} {\bibinfo {author} {\bibfnamefont {A.}~\bibnamefont
  {Spinelli}}, \bibinfo {author} {\bibfnamefont {M.~A.}\ \bibnamefont
  {Torija}}, \bibinfo {author} {\bibfnamefont {C.}~\bibnamefont {Liu}},
  \bibinfo {author} {\bibfnamefont {C.}~\bibnamefont {Jan}}, \ and\ \bibinfo
  {author} {\bibfnamefont {C.}~\bibnamefont {Leighton}},\ }\href@noop {}
  {\bibfield  {journal} {\bibinfo  {journal} {Phys. Rev. B}\ }\textbf {\bibinfo
  {volume} {81}},\ \bibinfo {pages} {155110} (\bibinfo {year}
  {2010})}\BibitemShut {NoStop}%
\bibitem [{\citenamefont {Mikheev}\ \emph {et~al.}(2015)\citenamefont
  {Mikheev}, \citenamefont {Hwang}, \citenamefont {Kajdos}, \citenamefont
  {Hauser},\ and\ \citenamefont {Stemmer}}]{mikheev2015}%
  \BibitemOpen
  \bibfield  {author} {\bibinfo {author} {\bibfnamefont {E.}~\bibnamefont
  {Mikheev}}, \bibinfo {author} {\bibfnamefont {J.}~\bibnamefont {Hwang}},
  \bibinfo {author} {\bibfnamefont {A.~P.}\ \bibnamefont {Kajdos}}, \bibinfo
  {author} {\bibfnamefont {A.~J.}\ \bibnamefont {Hauser}}, \ and\ \bibinfo
  {author} {\bibfnamefont {S.}~\bibnamefont {Stemmer}},\ }\href@noop {}
  {\bibfield  {journal} {\bibinfo  {journal} {Sci. Reports}\ }\textbf {\bibinfo
  {volume} {5}},\ \bibinfo {pages} {11079} (\bibinfo {year}
  {2015})}\BibitemShut {NoStop}%
\bibitem [{\citenamefont {Acevedo}\ \emph {et~al.}(2016)\citenamefont
  {Acevedo}, \citenamefont {Rubi}, \citenamefont {Lecourt}, \citenamefont
  {Lüders}, \citenamefont {Gomez-Marlasca}, \citenamefont {Granell},
  \citenamefont {Golmar},\ and\ \citenamefont {Levy}}]{acevedo2016}%
  \BibitemOpen
  \bibfield  {author} {\bibinfo {author} {\bibfnamefont {W.~R.}\ \bibnamefont
  {Acevedo}}, \bibinfo {author} {\bibfnamefont {D.}~\bibnamefont {Rubi}},
  \bibinfo {author} {\bibfnamefont {J.}~\bibnamefont {Lecourt}}, \bibinfo
  {author} {\bibfnamefont {U.}~\bibnamefont {Lüders}}, \bibinfo {author}
  {\bibfnamefont {F.}~\bibnamefont {Gomez-Marlasca}}, \bibinfo {author}
  {\bibfnamefont {P.}~\bibnamefont {Granell}}, \bibinfo {author} {\bibfnamefont
  {F.}~\bibnamefont {Golmar}}, \ and\ \bibinfo {author} {\bibfnamefont
  {P.}~\bibnamefont {Levy}},\ }\href@noop {} {\bibfield  {journal} {\bibinfo
  {journal} {Physics Letters A}\ }\textbf {\bibinfo {volume} {380}},\ \bibinfo
  {pages} {2870} (\bibinfo {year} {2016})}\BibitemShut {NoStop}%
\bibitem [{\citenamefont {Rom\'an~Acevedo}\ \emph {et~al.}(2017)\citenamefont
  {Rom\'an~Acevedo}, \citenamefont {Acha}, \citenamefont {S\'anchez},
  \citenamefont {Levy},\ and\ \citenamefont {Rubi}}]{acevedo2017}%
  \BibitemOpen
  \bibfield  {author} {\bibinfo {author} {\bibfnamefont {W.}~\bibnamefont
  {Rom\'an~Acevedo}}, \bibinfo {author} {\bibfnamefont {C.}~\bibnamefont
  {Acha}}, \bibinfo {author} {\bibfnamefont {M.~J.}\ \bibnamefont {S\'anchez}},
  \bibinfo {author} {\bibfnamefont {P.}~\bibnamefont {Levy}}, \ and\ \bibinfo
  {author} {\bibfnamefont {D.}~\bibnamefont {Rubi}},\ }\href@noop {} {\bibfield
   {journal} {\bibinfo  {journal} {Applied Physics Letters}\ }\textbf {\bibinfo
  {volume} {110}},\ \bibinfo {pages} {053501} (\bibinfo {year}
  {2017})}\BibitemShut {NoStop}%
\bibitem [{not()}]{nota}%
  \BibitemOpen
  \href@noop {} {\bibinfo  {journal} {We notice that although no ordering of
  oxygen vacancies was inferred from neutron powder diffraction performed in
  reduced LSMCO powders \cite{agua2011}, we did observe this ordering by high
  resolution transmission electron microscopy in LSMCO films, where a reduced
  brownmillerite-like structure was detected \cite{roman2020}}\ }\BibitemShut
  {NoStop}%
\end{thebibliography}%

\end{document}